\documentclass[twocolumn, trackchanges]{aastex631}
\usepackage{amsmath}
\hypersetup{
	colorlinks	= true,
	linkcolor	= red,
	urlcolor	= cyan,
	citecolor	= blue
}

\usepackage{gensymb}
\usepackage{amssymb}
\usepackage{xcolor}
\definecolor{mygreen}{rgb}{0.1, 0.7, 0.2}
\definecolor{myorange}{rgb}{1,0.5,0}
\definecolor{myred}{rgb}{1,0,0}

\usepackage{lipsum}
\usepackage{multirow}
\usepackage{makecell}

\graphicspath{{./images/}}

\newcommand{\exotr}{\mbox{\textsc{ExoTR}}}
\newcommand{\lhs}{LHS~1140~b}

\received{March 20$^{th}$, 2024}
\accepted{May 22$^{nd}$,  2024}

\submitjournal{ApJL}

\shortauthors{Damiano et al.}

\usepackage{fancyhdr}
\begin{document}
    \title{\lhs\ is a potentially habitable water world}
	
	\correspondingauthor{Mario Damiano}
	\email{mario.damiano@jpl.nasa.gov}
	
	\author[0000-0002-1830-8260]{Mario Damiano}
	\affiliation{Jet Propulsion Laboratory, California Institute of Technology, Pasadena, CA 91109, USA}
    
    \author[0000-0003-3355-1223]{Aaron Bello-Arufe}
	\affiliation{Jet Propulsion Laboratory, California Institute of Technology, Pasadena, CA 91109, USA}
	
	\author[0000-0002-1551-2610]{Jeehyun Yang}
	\affiliation{Jet Propulsion Laboratory, California Institute of Technology, Pasadena, CA 91109, USA}

 	\author[0000-0003-2215-8485]{Renyu Hu}
	\affiliation{Jet Propulsion Laboratory, California Institute of Technology, Pasadena, CA 91109, USA}
	\affiliation{Division of Geological and Planetary Sciences, California Institute of Technology, Pasadena, CA 91125, USA}
	
	\begin{abstract}
    \lhs\ is a small planet orbiting in the habitable zone of its M4.5V dwarf host. Recent mass and radius constraints have indicated that it has either a thick H$_2$-rich atmosphere or substantial water by mass. Here we present a transmission spectrum of \lhs\ between 1.7 and 5.2 $\mu$m, obtained using the NIRSpec instrument on JWST. By combining spectral retrievals and self-consistent atmospheric models, we show that the transmission spectrum is inconsistent with H$_2$-rich atmospheres with varied size and metallicity, leaving a water world as the remaining scenario to explain the planet's low density. Specifically, a H$_2$-rich atmosphere would result in prominent spectral features of CH$_4$ or CO$_2$ on this planet, but they are not seen in the transmission spectrum. Instead, the data favors a high-mean-molecular-weight atmosphere (possibly N$_2$-dominated with H$_2$O and CO$_2$) with a modest confidence. Forming the planet by accreting C- and N-bearing ices could naturally give rise to a CO$_2$- or N$_2$-dominated atmosphere, and if the planet evolves to or has the climate-stabilizing mechanism to maintain a moderate-size CO$_2$/N$_2$-dominated atmosphere, the planet could have liquid-water oceans. Our models suggest CO$_2$ absorption features with an expected signal of 20 ppm at 4.2 $\mu$m. As the existence of an atmosphere on TRAPPIST-1 planets is uncertain, \lhs\ may well present the best current opportunity to detect and characterize a habitable world.
	\end{abstract}
	
	\keywords{Exoplanet atmospheric composition -- JWST data analysis -- Transmission spectroscopy -- Bayesian statistics}
	
	\section{Introduction} \label{sec:intro}

    The endeavor to search for and characterize potentially habitable planets is driving the exploration of the Universe. In the near term, it is recognized that M dwarfs represent an ideal environment to discover and characterize small and temperate planets \citep{seager2010exoplanet}. The relatively small size of the host star makes it easier to detect atmospheric absorption and measure the planetary mass using the radial velocity technique. The lower irradiance levels of M dwarfs mean that the habitable zone (HZ), i.e., the range of orbits within which a planetary surface can support liquid water, is closer to the star. Planets in the HZ around M dwarfs have correspondingly short orbital periods and more frequent transits, making them more accessible for detailed characterization. 
    
    The only known Earth-sized ($<1.5$ R$_{\oplus}$) HZ planets potentially suitable for atmospheric studies by JWST are in the TRAPPIST-1 system \citep{gillon2017seven}. Ongoing TESS and ground-based planet surveys are expected to discover only $\sim1$ more temperate and Earth-sized planets like the TRAPPIST-1 planets \citep{kunimoto2022tess,sebastian2021tess}. Meanwhile, several temperate planets with radii in the $1.5-2.2\ R_{\oplus}$ range have been found and most of these larger-than-Earth planets are volatile-rich \citep{rogers2015most}. The exoplanet demographics \citep[e.g.,][]{luque2022density,rogers2023conclusive} and planet formation models \citep[e.g.,][]{venturini2020nature,izidoro2022exoplanet,chakrabarty2023water} indicate that they can have massive H$_2$-rich envelopes or a large fraction of water by mass (i.e., water worlds). The temperate sub-Neptunes located farther from their host stars are more likely to be water worlds \citep{izidoro2022exoplanet,chakrabarty2023water}. The possibility that some of the sub-Neptunes are water worlds is important, because temperate water worlds that have moderate-size atmospheres can host liquid-water oceans, and are thus targets for the search of habitability \citep{goldblatt2015habitability,koll2019hot,madhusudhan2021habitability}.
    
    However, the required conditions for a liquid water ocean surface on water worlds are likely more stringent than previously thought. Recent planetary climate models with self-consistent treatments of water vapor and cloud feedback indicate that water worlds orbiting M dwarfs would already enter the runaway greenhouse state if they receive $>\sim0.3\times$ Earth's insolation \citep{innes2023runaway,leconte20243d}. This requirement makes it much less plausible for warmer sub-Neptunes (K2-18~b and TOI-270~d, for example) to host liquid-water oceans unless they have fine-tuned conditions such as a very high Bond albedo. Therefore, the search for potentially habitable planets around M dwarfs must focus on even cooler planets.
    
    LHS 1140 is an M-dwarf with a mass and radius approximately 15\% that of the Sun and a temperature of $\sim$3000 K \citep{dittmann2017temperate,lillo2020lhsmass}. It hosts two planets of very different natures. LHS 1140 c, the inner planet, is a warm super-Earth ($\sim2$ M$\oplus$ and $\sim1.2$ R$\oplus$) with an equilibrium temperature of $\sim420$ K \citep{ment2019lhs,cadieux2024lhsmass}. \lhs\ was the first to be discovered \citep{dittmann2017temperate}, and it only receives $\sim42\%$ irradiation from the star as Earth receives from the Sun, leading to a zero-albedo equilibrium temperature of $\sim220$ K and placing the planet well within the habitable zone, either as a rocky planet with an N$_2$-CO$_2$ atmosphere \citep{kopparapu2013habitable}, or, with a modest Bond albedo of 0.3, as a water world with an H$_2$-rich atmosphere \citep{innes2023runaway,leconte20243d}.
        
    The physical properties of \lhs\ have been studied with high-precision transit and radial-velocity measurements (Table~\ref{tab:param}). According to \cite{lillo2020lhsmass}, \lhs\ would have a mass of $6.4\pm0.5$ M$\oplus$ and a radius of $1.64\pm0.05$ R$\oplus$. In this case, the mass and radius would be fully consistent with an Earth-like bulk composition, and detailed internal structure modeling suggests that the planet is likely iron-enriched but could also have an ocean more massive than Earth’s ocean \citep{lillo2020lhsmass}. However, recent measurements of \cite{cadieux2024lhsmass} suggest that the planet has a lower mass of $5.60\pm0.19$ M$\oplus$ and a larger radius of $1.73\pm0.025$ R$\oplus$. The refined constraints on the mass and radius indicate that the planet is substantially less dense than an Earth-like composition. Aside from the unlikely scenario of a coreless planet, the planet's low density can be explained by including $\sim0.1\%$ H$_2$/He or $10-20\%$ water by mass \citep{rogers2023conclusive,cadieux2024lhsmass}. Note that even with $\sim0.1\%$ H$_2$/He, the surface pressure of the envelope would be $>4000$ bar -- a ``massive'' envelope for atmospheric studies. The combination of the low temperature and lower-than-Earth-composition density makes \lhs\ a unique, naturally plausible, water world candidate for the observation of extant habitability.    

    It is thus crucial to determine whether \lhs\ has an H$_2$-rich envelope or a water-dominated one. With the expected thermal emission signal $<10$ ppm at 5 $\mu$m, transmission spectroscopy is the only feasible way to provide constraints on its atmospheric composition. An initial effort to characterize the atmosphere of \lhs\ was made by observing the planet with the Hubble Space Telescope (HST). Two HST/WFC3-G141 visits (centered at 1.3 $\mu$m) have been obtained for the transit of \lhs. The transmission spectra resulting from these observations suggested modulations that peak at 1.38 $\mu$m, apparently compatible with H$_2$O absorption in an H$_2$-dominated atmosphere \citep{edwards2020lhshst}. 
    It is however challenging to determine whether the spectral modulation is produced by the planetary atmosphere or by stellar heterogeneities  \citep[i.e., the transit light source effect,][]{rackham2018transit, moran2023high,lim2023atmospheric,may2023double}.

    We have observed two transits of \lhs\ using JWST (program ID: 2334, PI: M. Damiano) with the NIRSpec instrument, combining the G235H and G395H gratings to provide a wavelength coverage between $\sim1.7$ and $\sim5.2$ $\mu$m. Here we present the data analysis of the observations and the interpretation of the resulting transmission spectrum through Bayesian retrieval analysis and self-consistent atmospheric modeling. The manuscript is organized as follows: in Sec.~\ref{sec:method}, we will describe the observations, the data analysis procedure, the retrieval setup, and the atmospheric models. In Sec.~\ref{sec:result}, we will present the results of our analysis, including the corrected white and spectroscopic light curves, the extracted transmission spectrum, and the results from the retrieval and forward-model analyses. In Sec.~\ref{sec:discussion}, we will discuss the nature of the planet in light of these new observations and analyses and describe possible future observations to further unveil the planet's nature. We will conclude with Sec.~\ref{sec:conclusion} and summarize our findings of \lhs, which appears to be one of the best candidates for habitability studies today.
    
        \begin{deluxetable}{lcc}
            \tablecaption{System parameters used in this paper. (1) \cite{cadieux2024lhsmass}, (2) \cite{dittmann2017temperate}. \label{tab:param}}
            \tablehead{\textbf{Parameter} & \textbf{Value} & \textbf{Reference}}
            \startdata
                LHS 1140 & & \\
                M$_{\star}$ [M$_{\odot}$] & 0.1844$\pm$0.0045 & (1) \\
                R$_{\star}$ [R$_{\odot}$] & 0.2159$\pm$0.0030 & (1)\\
                $\rho_{\star}$ [g cm$^{-3}$] & 25.8$\pm$1.0 & (1)\\
                T$_{eff}$ [K] & 3096$\pm$48 & (1) \\
                L$_{\star}$ [L$_{\odot}$] & 0.0038$\pm$0.0003 & (1) \\
                SpT & M4.5V & (2) \\
                $[$Fe/H$]$ [dex]  & -0.15$\pm$0.09 & (1) \\
                log g [cgs] & 5.041$\pm$0.016 & (1) \\
                \hline
                \lhs & & \\
                P [days] & 24.73723$\pm$0.00002 & (1) \\
                t0 [BJD–2457000] & 1399.9300$\pm$0.0003 & (1) \\
                a [au] & 0.0946$\pm$0.0017 & (1)\\ 
                $i$ [deg] & 89.86$\pm$0.04 & (1) \\
                R$_p$ [R$_{\oplus}$] & 1.730$\pm$0.025 & (1) \\
                M$_p$ [M$_{\oplus}$] & 5.60$\pm$0.19 & (1) \\
                $\rho$ [g cm$^{-3}$] & 5.9$\pm$0.3 & (1) \\
                T$_{eq}$ [K] & 226$\pm$4 & (1) \\
            \enddata
        \end{deluxetable}
	
	\section{Methods} \label{sec:method}

    \subsection{Observations} \label{sec:obs}

    Two primary transits of \lhs\ have been observed by JWST on July 5$^{th}$ and July 30$^{th}$, 2023 and the datasets are available in the MAST archive (dataset: \dataset[10.17909/r627-v590]{http://dx.doi.org/10.17909/r627-v590}).
    The two visits were recorded with the NIRSpec instrument, using the G235H and G395H gratings to cover a wide wavelength range from 1.665 $\mu$m to 5.175 $\mu$m. We used the same slit, sub-array, and readout pattern for both observations. Different exposure times per integration were used for the two observations to yield a similar maximum saturation. The numeric details of the two observations are reported in Table \ref{tab:obs}. The two visits consist of the in-transit event ($\sim$117 minutes), and a $1.5\times$ such amount of time out of transit ($\sim$157 minutes before and $\sim$40 minutes after the transit).

        \begin{deluxetable}{lcc}
            \tablecaption{Observation details. \label{tab:obs}}
            \tablehead{ & \textbf{G235H} & \textbf{G395H}}
            \startdata
                Slit & S1600A1 & S1600A1 \\
                Sub-array & SUB2048 & SUB2048 \\
                Readout pattern & NRSRAPID & NRSRAPID \\
                n. groups & 8 & 15 \\
                n. integrations & 2382 & 1342\\
                Exp. time per integration [sec] & 8.14 & 14.45\\
                Total exp. time [sec] & 19385.86 & 19391.90 \\
                Total exp. time [hrs] & 5.385 & 5.387 \\
            \enddata
        \end{deluxetable}
    

    \subsection{Data analysis} \label{sec:analysis}
   
    We reduced the NIRSpec data using \texttt{Eureka!} \citep[version 0.10,][]{bell2022eureka}, an open-source end-to-end pipeline for exoplanet time-series observations (TSO). Recent works have demonstrated the ability of \texttt{Eureka!} to produce spectra that are consistent with other state-of-the-art JWST pipelines when applied to NIRSpec/G395H data \citep[e.g.][]{alderson2023wasp39,lustigyaeger2023lhs475} and to other JWST instrument modes \citep{ers2023wasp39,ahrer2023wasp39}. During the data reduction and light curve fits, we treated the data from the two gratings (G235H and G395H) and the two detectors (NRS1 and NRS2) independently. 
    
    Starting from the \textit{uncal.fits} files, we ran stages 1 and 2 of \texttt{Eureka!} to process and calibrate the raw data. After testing different setups, we decided to run all the default stage 1 near-infrared TSO steps. We corrected the super-bias using a scale factor calculated with background pixels located at least 8 pixels away from the trace. We applied a smoothing filter of length 62 integrations to the scale factor values. We set the jump rejection threshold to $5\sigma$ and ran a group-level background subtraction using the average of the background pixels in each detector column. In stage 2, we skipped the flat field step, which increases the noise in the data, and the photometric calibration step, as we are only interested in the relative flux measurements. 
    
\begin{figure*}
            \centering
            \includegraphics[width=1\linewidth]{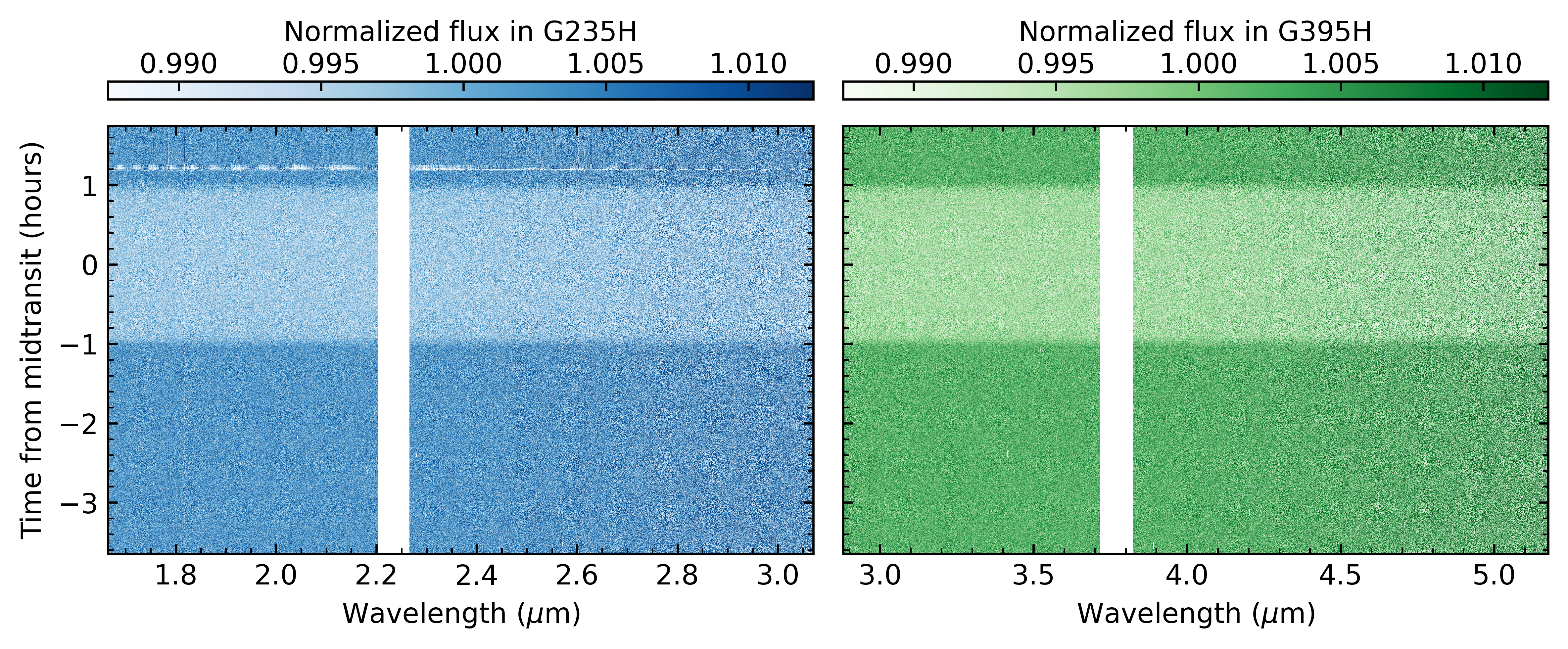}
            \caption{Raw spectroscopic light curves from the NIRSpec/G235H (left) and G395H (right) modes. Both modes show a gap that separates the data from the NRS1 and NRS2 detectors. The white horizontal stripe in the G235H data are points affected by the high-gain antenna move.}
            \label{fig:speclcs}
        \end{figure*}

    Stages 3 and 4 of \texttt{Eureka!} perform optimal spectral extraction and generate the spectroscopic light curves. We extracted columns in the range 545--2041 and 6--2044 for the NRS1 and NRS2 data, respectively. We corrected the curvature of the trace, and we performed a second background subtraction using the average value in each detector column of pixels located at least 9 pixels away from the trace. For the optimal extraction of the spectra, we constructed the spatial profile from the median frame, and we used an aperture region with a half width of 3 pixels. We generated spectroscopic light curves at the native pixel resolution, spanning the following regions: 1.665-2.202~$\mu$m (G235H/NRS1), 2.266-3.070~$\mu$m (G235H/NRS2), 2.880-3.717~$\mu$m (G395H/NRS1), 3.824-5.175~$\mu$m (G395H/NRS2). By native pixel resolution, we refer to the extraction of light curves at the individual pixel column level. We analyzed every column of the detector, and the spectral resolution is Nyquist-sampled by two columns per resolution element. The extracted light curves are shown in Fig.~\ref{fig:speclcs}.
        
\begin{figure*}
            \centering
            \includegraphics[width=\linewidth]{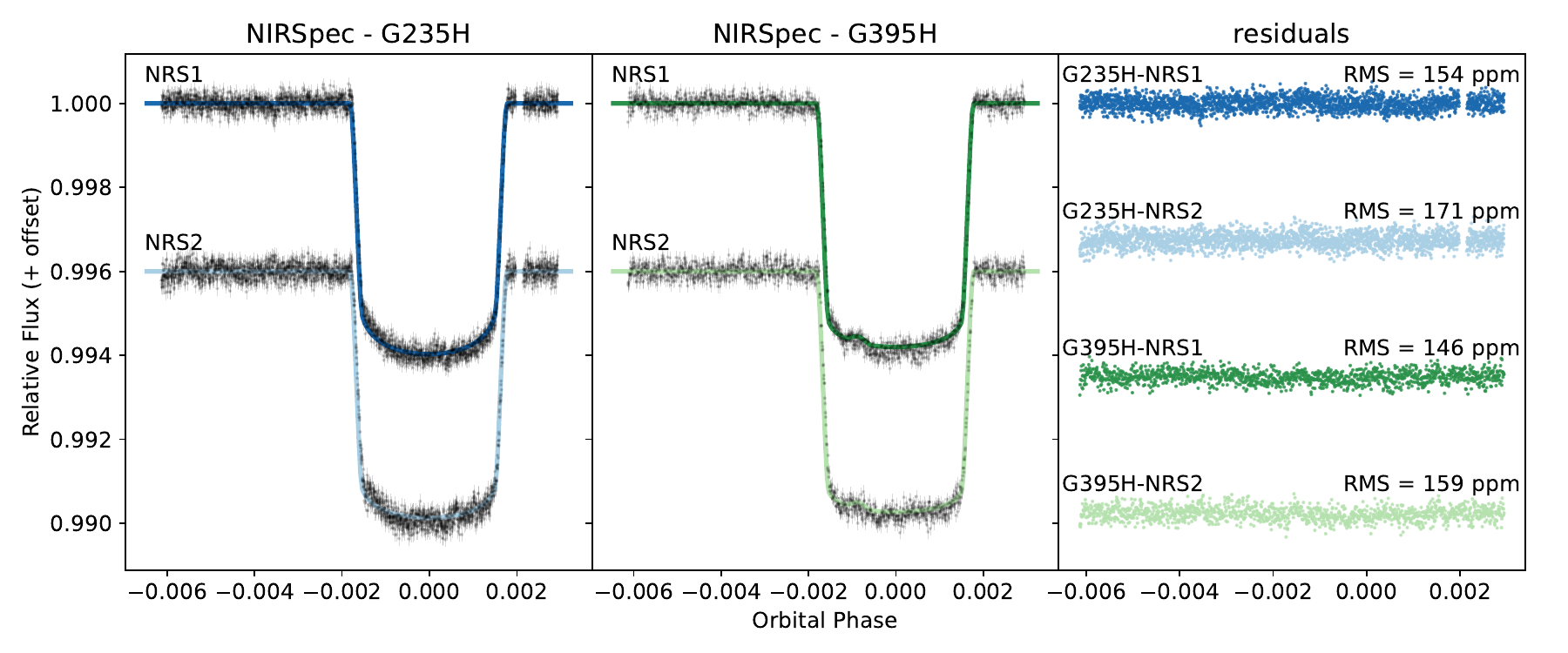}
            \caption{White light curves analysis of the two \lhs\ transits. The white light curve of each visit and each detector is normalized. \textbf{Left panel} shows the white light curves from the NRS1 and NRS2 detectors using the G235H grating. \textbf{Central panel} shows the same information as the left but for the G395H grating. The starspot crossing can be seen here. \textbf{Right panel} shows the residuals of the fit for each of the four white light curves. The relative root mean square (RMS) is also noted. \label{fig:wlc_analysis}}
        \end{figure*}

    We modeled the light curves as a combination of a \texttt{batman} transit function \citep{kreidberg2015batman} and a linear polynomial in time. First, we fitted each white light curve to determine the transit midpoint ($T_0$), orbital inclination ($i$), and scaled semi-major axis ($a/R_{\star}$). During the white light curve fits, we kept the quadratic limb darkening coefficients free. We included a white-noise multiplier as a free parameter in the fits. We masked the integrations in the range 2134--2185 of the G235H light curves, which are affected by a high-gain antenna move (Fig.~\ref{fig:speclcs}). Meanwhile, the G395H white light curves show a starspot crossing event, visible in both detectors (Fig.~\ref{fig:wlc_analysis}). 
        
    
    We corrected the white light curves affected by the starspot using the semi-analytical spot modeling code \texttt{spotrod} \citep{beky2014spotrod}. The spot is characterized by four parameters in this model: the ratio of the spot's radius to the star's radius (R$_{spot}$/R$_{\star}$), the ratio of the spot's intensity compared to the unspotted surface of the star ($f$), and the position of the spot's center on the star's surface, represented by two coordinates ($\theta$, $r^2$). The spot-corrected white light curves are then fitted to obtain the photometric $R_p/R_{star}$, $T_0$, $i$, and $a/R_{star}$ for each of the four datasets (Table~\ref{tab:wlc_result}).
        
    We then fitted the spectroscopic light curves at the native resolution, keeping $T_0$, $i$, and $a/R_{star}$ fixed to the values derived from the white light curves. We masked the integrations affected by the high-gain antenna move and the star-spot crossing event and fixed the quadratic limb darkening coefficients to those calculated by the \texttt{ExoTiC-LD} package \citep{grant2022exotic} using 3D stellar models \citep{magic2015stagger} and assuming the stellar parameters as those reported by \citet{cadieux2024lhsmass}. As in the white light curve fits, we included a white-noise multiplier to ensure that the uncertainties of the results are consistent with the scatter of the residuals.  We also tried fitting the spectroscopic light curves without masking the spot crossing event and found that the resulting transmission spectra were consistent.

	\subsection{Atmospheric retrieval setup} \label{sec:retrieval}

    We used the \texttt{ExoTR} \footnote{\textbf{Publicly available on GitHub: \href{https://github.com/MDamiano/ExoTR}{ExoTR}}} (Exoplanetary Transmission Retrieval) algorithm to interpret the derived transmission spectrum. \texttt{ExoTR} is a fully Bayesian retrieval algorithm designed to interpret exoplanet transmission spectra. Its useful features include: a) the cloud layer can be modeled as an optically thick surface or as a physically motivated cloud scenario tied to a non-uniform water volume mixing ratio profile, similarly to \texttt{ExoReL$^\Re$} \citep{hu2019information,damiano2020exorel,damiano2022small}, b) the stellar heterogeneity components can be jointly fit with the planetary atmospheric parameters \citep{rackham2017heterogeneity,pinhas2018retrieval}, c) the atmospheric abundances are fit in the centered-log-ratio (CLR) space and the prior functions are designed to render a flat prior when transformed back to the log-mixing-ratio space \citep{damiano2021prior}, and d) the possibility to fit photochemical hazes with prescribed optical constants and a free particle size. \texttt{ExoTR} will be described in detail in a subsequent paper (Tokadjian et al. in prep.).

        \begin{deluxetable*}{l|cc|cc}
            \tablecaption{White light curves analysis results. \label{tab:wlc_result}}
            \tablehead{\textbf{Parameter} & \textbf{G235H -- NRS1} & \textbf{G235H -- NRS2} & \textbf{G395H -- NRS1} & \textbf{G395H -- NRS2}}
            \startdata
                T$_0$ [BJD] & 60131.037574 $\pm$ 0.000016 & 60131.037605 $\pm$ 0.000018 & 60155.774828 $\pm$ 0.000020& 60155.774826 $\pm$ 0.000022 \\
                R$_p$/R$_{\star}$ & 0.07422$_{-0.00008}^{+0.00011}$ & 0.07437$_{-0.00012}^{+0.00012}$ & 0.07404$_{-0.00012}^{+0.00013}$ & 0.07441$_{-0.00013}^{+0.00012}$ \\
                a/R$_{\star}$ & 95.6$_{-0.7}^{+0.5}$ & 94.4$_{-0.7}^{+0.7}$ & 95.1$_{-0.8}^{+0.7}$ & 91.6$_{-0.8}^{+0.9}$\\
                $i$ [deg] & 89.93$_{-0.03}^{+0.03}$ & 89.88$_{-0.02}^{+0.03}$ & 89.92$_{-0.03}^{+0.04}$ & 89.81$_{-0.02}^{+0.02}$ \\
                $u$1 & 0.145$_{-0.017}^{+0.017}$ & 0.151$_{-0.020}^{+0.022}$ & 0.091$_{-0.022}^{+0.022}$ & 0.012$_{-0.009}^{+0.016}$ \\
                $u$2 & 0.223$_{-0.032}^{+0.032}$ & 0.152$_{-0.039}^{+0.038}$ & 0.250$_{-0.042}^{+0.040}$ & 0.247$_{-0.030}^{+0.025}$ \\
            \enddata
        \end{deluxetable*}
        
    In this work, we included the offset between the NRS1 and NRS2 detectors for both gratings as free parameters ($off_{n}$), given recent results using NIRSpec/G395H \citep[e.g.,][]{moran2023high,madhusudhan2023k218}. We kept the NIRSpec/G235H-NRS1 dataset fixed and applied the offsets to other datasets. $off_1$ is the offset between NRS1 and NRS2 within the G235H grating, $off_2$ is the offset between G325H-NRS1 and G395H-NRS1, and $off_3$ is the offset between G325H-NRS1 and G395H-NRS2. Here, we adopted a simple cloud model characterized by the cloud top pressure (${\rm P}_{top}$) and kept the planetary temperature at 200 K. The atmospheric abundances are parameterized in the CLR space with H$_2$ or N$_2$ as the background gasses, and the other gases considered as free parameters in the retrieval include H$_2$O, CH$_4$, NH$_3$, CO$_2$, and CO. We jointly fit the impact of stellar heterogeneity with the planetary parameters. Three parameters were used to describe the stellar component \citep{pinhas2018retrieval}: the star temperature ($Ts_{phot}$), the star heterogeneity temperature (either faculae or spots) ($Ts_{het}$), and the fraction of stellar surface impacted by the heterogeneity ($\delta$). The stellar spectra are adopted from the PHOENIX models \citep{husser2013phoenix}. Both the modeled transmission spectra and stellar models are calculated at the data spectral resolution and wavelength bins. The spectral resolution-linked bias (RLB) effect \citep{deming2017rlb} is known to bias planetary transmission spectra if the star is not modeled at high spectral resolution so that individual molecular absorption lines are resolved. In \cite{deming2017rlb}, the RLB effect is shown to have a direct correlation with the atmospheric scale height. In the case of \lhs, the low equilibrium temperature and potentially high mean molecular mass inversely correlate with the RLB effect, and so we do not expect a significant impact on the results presented in this work.
    
    Table~\ref{tab:retrieval_setup} lists the free parameters, the prior space used, and the range in which the parameters are probed. \texttt{ExoTR} uses \texttt{MultiNest} \citep{feroz2009multinest} to sample the Bayesian evidence, estimate the parameters, and determine the posterior distribution functions. \texttt{MultiNest} is used through its \texttt{Python} implementation \texttt{pymultinest} \citep{buchner2014multinest}. For all the retrieval analyses presented here, we used 500 live points and 0.5 as the Bayesian evidence tolerance. Finally, to assess the significance of a scenario over the null hypothesis, we calculated the Bayes factor \citep{trotta2008bayes}, which is a quantitative statistical measurement to choose one model over another one.
    
        \begin{deluxetable*}{lcl}
            \tablecaption{Model parameters and prior probability distributions used in the atmospheric retrievals. $\mathcal{U}(a,b)$ is the uniform distribution between values $a$ and $b$, $\mathcal{LU}(a,b)$ is the log-uniform (Jeffreys) distribution between values $a$ and $b$, and $\mathcal{N}(\mu,\sigma^2)$ is the normal distribution with mean $\mu$ and variance $\sigma^2$. NOTE - $(^1)$ \cite{damiano2021prior}, $(^2)$ \cite{cadieux2024lhsmass}. \label{tab:retrieval_setup}}
            \tablehead{\textbf{Parameter} & \textbf{Symbol} & \textbf{Prior}}
            \startdata
                Datasets offsets [ppm] & $off_n$ & $\mathcal{U}$(-100, 100) \\
                Planetary radius [R$_{\oplus}$] & $R_p$ & $\mathcal{U}$(0.5, 2)$\times$ R$_p(^2)$ \\
                Cloud top [Pa] & P$_{top}$ & $\mathcal{LU}$(0.0, 9.0)\\
                VMR H$_2$O & H$_2$O & CLR$(-25.0, 25.0)(^1)$ \\
                VMR CH$_4$ & CH$_4$ & CLR$(-25.0, 25.0)(^1)$  \\
                VMR NH$_3$ & NH$_3$ & CLR$(-25.0, 25.0)(^1)$  \\
                VMR CO$_2$ & CO$_2$ & CLR$(-25.0, 25.0)(^1)$  \\
                VMR CO & CO & CLR$(-25.0, 25.0)(^1)$  \\
                VMR N$_2$ & N$_2$ & CLR$(-25.0, 25.0)(^1)$  \\
                Heterogeneity fraction & $\delta$ & $\mathcal{U}$(0.0 - 0.5) \\
                Heterogeneity temperature [K] & $Ts_{het}$ & $\mathcal{U}$(0.5, 1.2) $\times$ $Ts_{phot}(^2)$\\
                Stellar temperature [K] & $Ts_{phot}$ & $\mathcal{N}$(3096, 48)$(^2)$ \\
            \enddata
        \end{deluxetable*}

    \subsection{Self-consistent atmospheric models} \label{sec:forward}

    We also simulated representative self-consistent atmospheric models for the potential scenarios of \lhs. We explored massive H$_2$-rich atmospheres with $1\times$, $10\times$, and $100\times$ solar metallicities, a small H$_2$-dominated atmosphere with CO$_2$ \citep[i.e., a potential ``hycean'' scenario following][]{madhusudhan2021habitability,hu2021unveiling}, as well as N$_2$- and CO$_2$-dominated atmospheres, which are plausible atmospheres overlaying a water-dominated mantle (see Sec.~\ref{sec:ww}).
        
    We calculated the pressure-temperature profiles for each scenario under radiative-convective equilibrium using the climate module of the ExoPlanet Atmospheric Chemistry \& Radiative Interaction Simulator (EPACRIS-Climate, Scheucher et al. in prep.). The model solves the radiative fluxes using the 2-stream formulation of \cite{heng2018radiative} and performs the moist adiabatic adjustment using the formulation of \cite{graham2021multispecies}. Water is treated as condensable in these models and its atmospheric abundance is self-consistently adjusted together with the moist adiabats. For simplicity, we did not include the cloud albedo feedback in these calculations, but instead applied a modest albedo of 0.3 in all models.
        
    Based on the pressure-temperature profiles, we then simulated the effects of vertical transport and photochemistry using the chemistry module of EPACRIS. For the massive atmosphere models, we used a chemical network generated by the Reaction Mechanism Generator \citep{gao2016rmg,johnson2022rmg} for the conditions relevant to \lhs\ and coupled it to the planetary-scale kinetic-transport model \citep{yang2024epacris}. For the small atmosphere models, we used the chemical network in \cite{hu2021unveiling} together with the recent rate updates from \cite{wogan2024k218b}. The resulting chemical abundance profiles were used together with the pressure-temperature profiles to calculate the transmission spectra. The impact of water condensation and cloud formation is self-consistently included in the transmission spectra such that the cold trap in the atmosphere controls the pressure level of the cloud and the water vapor mixing ratio above the cloud \citep{hu2019information,damiano2020exorel}.

	\section{Results} \label{sec:result}
    
    \subsection{White light curves and transmission spectrum}
            
    Fig.~\ref{fig:speclcs} depicts the extracted and calibrated light curves at the native resolution. In the light curves obtained with the G235H grating, we observed the high-gain antenna move effect. We masked the data affected by this distortion. The high-gain antenna move happened outside the transit, and therefore masking these data does not result in any significant loss of signal.


    The signal extracted from the four detectors resulted in four white light curves. Fig.~\ref{fig:wlc_analysis} shows the analysis performed on each of the four white light curves as well as the residuals. The white light curves obtained with the G395H grating show the effect of a starspot crossing at the orbital phase of $\sim0.0015$ earlier than the mid-transit point. First, we fit a model without the starspot using \texttt{spotrod} \citep{beky2014spotrod} and \texttt{MultiNest} \citep{feroz2009multinest,espinoza2019starspot} combined, so that we can calculate the Bayesian evidence of the model. Then, we fit a model that includes the starspot distortion. By calculating the Bayes factor, we found that the model with one starspot is preferred by the data with a significance greater than 5$\sigma$ over the light curve model without starspot. For the one starspot model, we obtained an intensity ratio of $f=0.963\pm0.012$ and a spot-to-star radius ratio of R$_{spot}$/R$_{\star}=0.176\pm0.041$. The white light curve fitting results are summarized in Table~\ref{tab:wlc_result}. 
            
\begin{figure*}
                \centering
                \includegraphics[width=\linewidth]{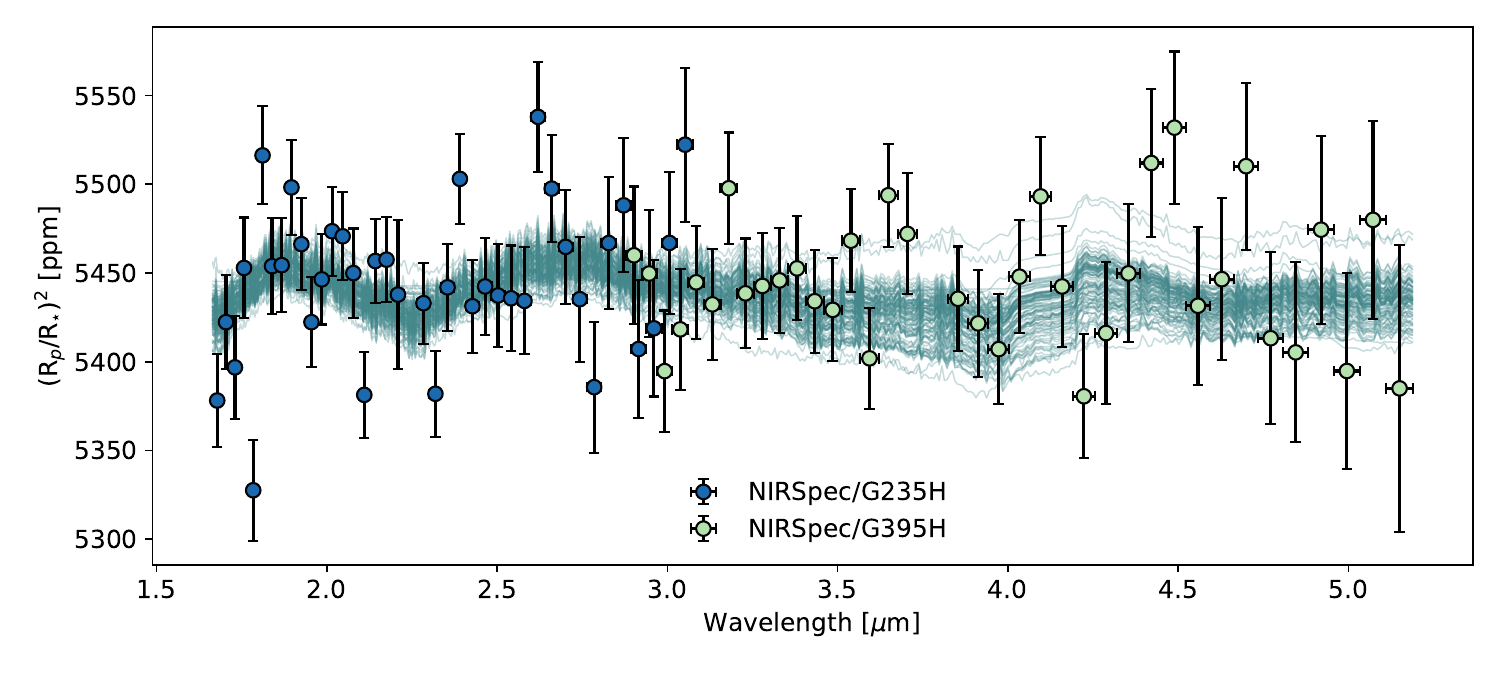}
                \caption{Transmission spectrum of \lhs. The two transit observations are color coded, i.e., blue for the G235H grating and green for the G395H grating. The models shown are randomly selected solutions from the posterior distributions of Scenario 4 (Table~\ref{tab:retrieval_results}, corresponding to N$_2$-H$_2$O atmospheres with deep clouds). We have applied the best-fit offsets to the data. The offset between the two visits is $\sim50$ ppm, while the offset between the two detectors within each of the two visits is negligible (see Figure \ref{fig:1D_post} and Table~\ref{tab:retrieval_values_phys}).  \label{fig:spec_result}}
            \end{figure*}

    The planetary and transit parameters derived from the white light curves are then fixed and used to fit the spectroscopic light curves. We did not use any binning when fitting the spectroscopic light curves, in either the time domain or the wavelength domain. After obtaining the transmission spectrum at the native pixel resolution (R$\sim$2700), we binned the spectrum to a spectral resolution of R=65 to reveal any major molecular absorption features. The derived transmission spectrum is shown in Fig.~\ref{fig:spec_result}.

    \subsection{Retrieval results}

    We used \texttt{ExoTR} described in Sec.~\ref{sec:retrieval} to interpret the transmission spectrum of \lhs. We fixed the planetary and stellar parameters to those in Table~\ref{tab:param} and fit the atmospheric parameters in Table~\ref{tab:retrieval_setup}. The scenarios considered and the Bayesian evidence obtained are summarized in Table~\ref{tab:retrieval_results}, and the constraints of parameters are reported in Tables~\ref{tab:retrieval_values_phys} and \ref{tab:retrieval_values_chem} in Appendix \ref{sec:appendixA}.
            
    We considered two baseline scenarios and a bare rock scenario as the null hypotheses. The first of the two baseline scenarios assumes an H$_2$-only atmosphere (with Rayleigh scattering and H$_2$-H$_2$ collision-induced absorption), a generic cloud top, and the offsets between datasets. The other baseline scenario instead assumes an N$_2$-only atmosphere (with Rayleigh scattering) also with a generic cloud and offsets. The bare rock scenario only has the planetary radius and offsets as free parameters. In this way, we define the null hypotheses with different atmospheric scale heights or no atmosphere altogether. We ran retrievals on the data for these scenarios and obtained log-Bayesian evidence of ln(EV)$=$642.59 and 643.61 for the two baseline scenarios, respectively, and ln(EV)$=$643.16 for the bare rock scenario.
            
    On top of the baseline and bare rock scenarios, we successively considered various absorbing molecules to see which ones would be favored by the data. We added H$_2$O to either H$_2$- or N$_2$-dominated atmosphere and H$_2$O and CO$_2$ to N$_2$-dominated atmosphere. We also considered the general cases where H$_2$, N$_2$, H$_2$O, CO$_2$, CH$_4$, NH$_3$, and CO are all included in the retrieval. Table~\ref{tab:retrieval_results} ranks these scenarios by the order of decreasing Bayesian evidence. We found that the models with an N$_2$-dominated atmosphere with H$_2$O and CO$_2$ are favored over the baseline scenarios by $>3\sigma$ (Table~\ref{tab:retrieval_results}, Scenarios 1 and 2), while the inclusion of other gases does not result in the increase of evidence (Scenario 3). We also found that the Bayesian evidence does not change substantially when the stellar heterogeneity components are fit as free parameters in the retrieval (e.g., comparing Scenarios 4 and 1), and with stellar heterogeneity, the N$_2$-dominated atmosphere with H$_2$O and CO$_2$ is still preferred over the baseline scenarios by $\sim3\sigma$. 
    
\begin{figure*}
                \centering
                \includegraphics[width=\linewidth]{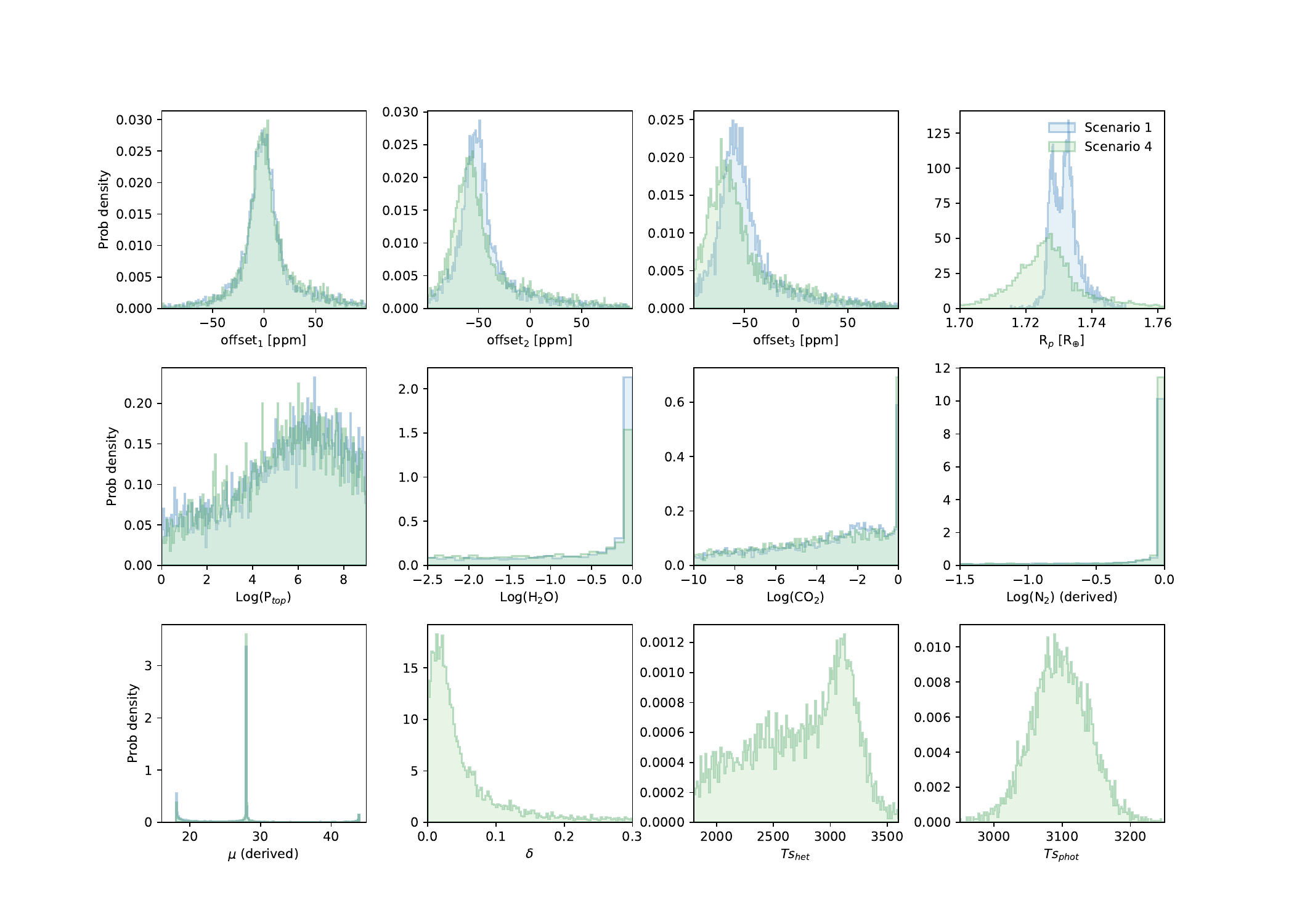}
                \caption{Comparison of the 1D posterior distribution functions of Scenarios 1 and 4 (i.e., including N$_2$, H$_2$O, CO$_2$, clouds and with and without the stellar heterogeneity). \label{fig:1D_post}}
            \end{figure*}

    The posterior distributions of Scenarios 1 and 4 are shown in Fig.~\ref{fig:1D_post}. While both retrievals suggest solutions that range from N$_2$- to H$_2$O-dominated atmospheres, the inclusion of the stellar heterogeneity components as free parameters decrease the probability density of high mixing ratios of H$_2$O, suggesting that the spectral modulation seen at wavelengths $<3\mu$m may be partly due to stellar heterogeneity. In both scenarios, low mixing ratios of H$_2$O are possible, making an N$_2$-dominated atmosphere more likely (also see Table~\ref{tab:retrieval_values_chem}.)
    
    Meanwhile, including H$_2$O in an H$_2$-dominated atmosphere results in an increase of evidence that corresponds to $\sim2\sigma$ with respect to the baseline scenarios (Scenario 6 in Table~\ref{tab:retrieval_results}). The cloud pressure required by this scenario is $\sim10^{2.8\pm0.5}$ Pa (Table~\ref{tab:retrieval_values_phys}). Based on the self-consistent atmospheric models, however, the water clouds should be located at a higher pressure on this planet, and an H$_2$-dominated atmosphere should have additional absorbing gas including CH$_4$ or CO$_2$ (see Sec.~\ref{sec:atmospheremodel}). Moreover, if we compare this scenario to an N$_2$-dominated atmosphere with H$_2$O (i.e., comparing Scenarios 6 and 2), the data still prefers the N$_2$-dominated atmosphere with a significance of $\sim2.8\sigma$ (Table~\ref{tab:retrieval_results}).

    We also performed a joint analysis of all available datasets including the transmission spectra obtained by HST and ground-based observations (Appendix~\ref{sec:appendixB}). In that case, the spectral contribution from stellar heterogeneity is clearly required to fit the dataset across the entire wavelength range covered. 

    Lastly, we performed an additional test to address how the data noise impacts the retrieval result. We simulated synthetic data by taking the value of the planetary radius (flat spectrum), adopting the same errorbars as the real data, and adding Gaussian noise to the data points based on the errorbar. We tried retrieving two scenarios (bare rock and atmosphere, same as scenario 3 and 8 in Table~\ref{tab:retrieval_results}) from this dataset and compared the resulted Bayesian evidence. We repeated this test for 10 different Gaussian noise realizations. For the majority of the tests (8 out 10), we observed that the bare rock scenario is preferred over the atmospheric scenario, suggesting that no evidence of atmosphere is present. Only in two cases, we observed that the Bayesian evidence of the atmospheric scenario was higher for the bare rock scenario. In both cases, the significance is $<2\sigma$. From the real data presented here, we found $>\sim3\sigma$ significance when the atmospheric scenario is compared to the bare rock scenario.
    
            \begin{deluxetable*}{p{0.45\textwidth} | c | cc}
                \tablecaption{Atmospheric scenarios considered and their corresponding log-Bayesian evidence. The third and fourth columns indicate the significance level at which each scenario is favored against the baseline scenarios, obtained by calculating the Bayes factor. The offsets between the visits and the detectors have been included as free parameters for all the scenarios (3 offsets). \label{tab:retrieval_results}}
                \tablehead{\textbf{Free parameters} & \textbf{log-Bayesian Evidence, ln(EV)} & \textbf{$\sigma$ baseline 1} & \textbf{$\sigma$ baseline 2}}
                \startdata
                    \textbf{1.} N$_2$, H$_2$O, CO$_2$, and cloud & 647.20 $\pm$ 0.16 & 3.48$\sigma$ & 3.18$\sigma$ \\
                    \hline
                    \textbf{2.} N$_2$, H$_2$O, and cloud & 646.68 $\pm$ 0.16 & 3.33$\sigma$ & 3.02$\sigma$ \\
                    \hline
                    \textbf{3.} H$_2$, N$_2$, H$_2$O, CH$_4$, NH$_3$, CO$_2$, CO, and cloud & 646.00 $\pm$ 0.16 & 3.12$\sigma$ & 2.67$\sigma$ \\
                    \hline
                    \textbf{4.} N$_2$, H$_2$O, CO$_2$, cloud, and stellar heterogeneity & 645.68 $\pm$ 0.16 & 3.03$\sigma$ & 2.59$\sigma$ \\ 
                    \hline
                    \textbf{5.} H$_2$, N$_2$, H$_2$O, CH$_4$, NH$_3$, CO$_2$, CO, cloud, and stellar heterogeneity & 645.22 $\pm$ 0.16 & 2.78$\sigma$ & 2.38$\sigma$ \\
                    \hline
                    \textbf{6.} H$_2$, H$_2$O, and cloud & 644.11 $\pm$ 0.17 & 2.34$\sigma$ & $<$2$\sigma$ \\
                    \hline
                    \textbf{7.} N$_2$-only atmosphere with cloud (i.e., baseline 2) & 643.61 $\pm$ 0.15 & 2.11$\sigma$ & $-$ \\
                    \hline
                    \textbf{8.} Bare rock & 643.16 $\pm$ 0.15 & $<$2$\sigma$ & $-$ \\
                    \hline
                    \textbf{9.} H$_2$-only atmosphere with cloud (i.e., baseline 1) & 642.59 $\pm$ 0.15 & $-$ & $-$ \\
                \enddata
            \end{deluxetable*}

    \subsection{Self-consistent atmospheric models}
    \label{sec:atmospheremodel}
            
\begin{figure}
                \centering
                \includegraphics[width=\columnwidth]{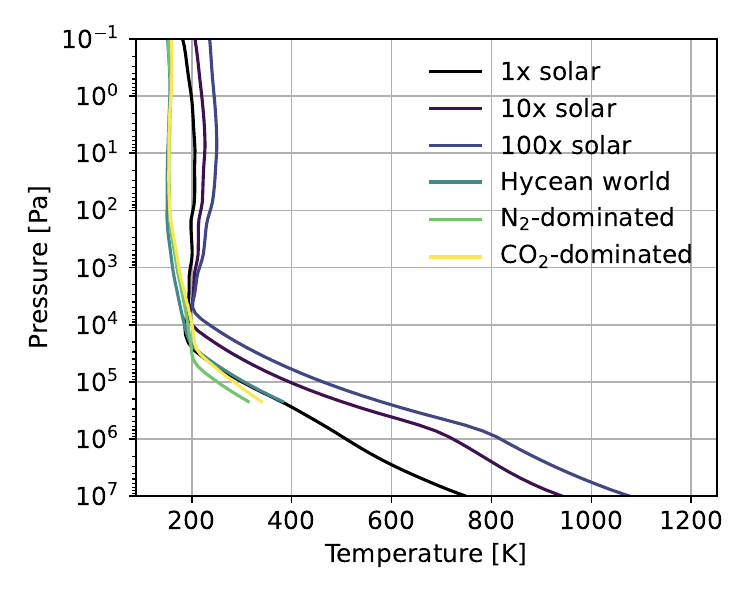}
                \caption{Modeled pressure-temperature profiles for \lhs, considering massive H$_2$-rich atmospheres with $1\times$, $10\times$, and $100\times$ solar metallicity abundances, and small (2.2-bar) H$_2$-, N$_2$-, and CO$_2$-dominated atmospheres. The small H$_2$-dominated atmosphere has 1\% CO$_2$, corresponding to the proposed ``hycean world'' scenario for temperate sub-Neptunes \citep{madhusudhan2021habitability,hu2021unveiling}. The temperature profiles indicate a cold trap at the pressure of $\sim10^4$ Pa for the H$_2$-rich atmospheres and $\sim10^{4.5}$ Pa for N$_2$- and CO$_2$-dominated atmospheres. The resulting transmission spectra of these models are shown in Fig.~\ref{fig:lhs_selfmod}.
                \label{fig:lhs_tp}}
            \end{figure}

\begin{figure*}
                \centering
                \includegraphics[width=\linewidth]{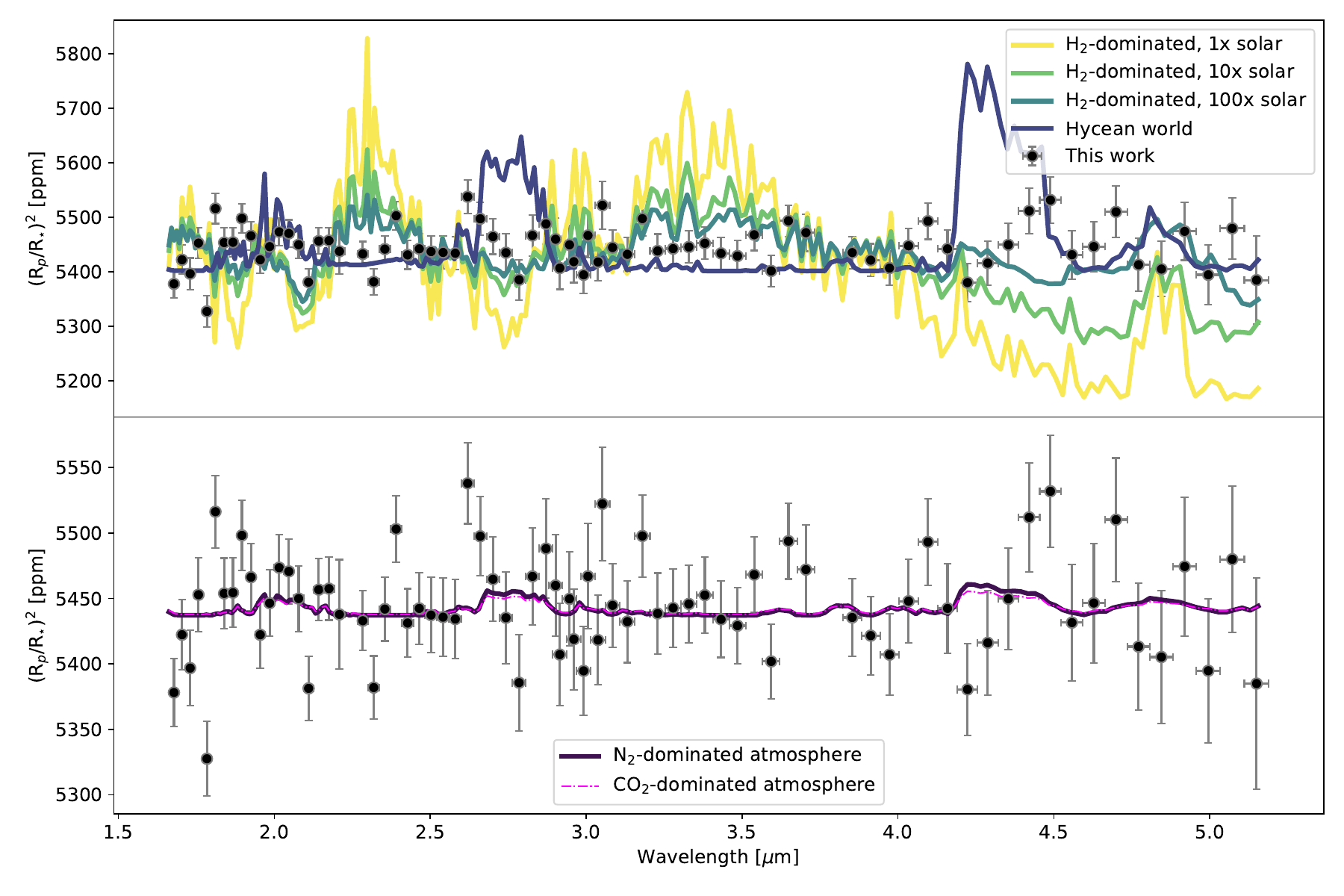}
                \caption{Self-consistent atmospheric models of \lhs\ compared to the JWST data. \textbf{Top panel} shows the massive H$_2$-rich atmosphere models with varied metallicities as well as the small H$_2$-dominated atmosphere with CO$_2$ model (``hycean world''). $\chi^2=$1406, 506, and 287 for the $1\times$, $10\times$, and $100\times$ solar metallicity models, respectively, and $\chi^2=$470 for the hycean model (considering 75 degrees of freedom). These models are ruled out by the data with a p-value $<0.00001$. \textbf{Bottom panel} shows the N$_2$- and CO$_2$-dominated atmosphere models. $\chi^2=$118 for both models.
            \label{fig:lhs_selfmod}}
            \end{figure*}

    Figure~\ref{fig:lhs_tp} shows the pressure-temperature profiles of our self-consistent atmosphere models. We find that, for H$_2$-rich atmospheres with a wide range of metallicity, the condensation of water extends to $\sim10^4$ Pa and the temperature at the cold trap would be $\sim200$ K. Meanwhile, the atmospheric chemistry models indicate that CH$_4$ and NH$_3$ should be the most abundant carbon and nitrogen species, with appreciable amount of CO2 occurring at the high metallicity of $100\times$ solar. This behavior is similar to the findings of \cite{hu2021photochemistry}. For \lhs, due to the low temperature, the mixing ratio of H$_2$O is reduced by $\sim3$ orders of magnitude above the cold trap pressure ($\sim10^4$ Pa). As a result, the transmission spectra of a massive H$_2$-rich atmosphere on \lhs, regardless of metallicity, would at least show strong spectral features of CH$_4$, which is clearly ruled out by the transmission spectrum measured here (Figure \ref{fig:lhs_selfmod}).

    If the planet has a small H$_2$-dominated atmosphere on top of an H$_2$O-dominated mantle, the dominant form of carbon should be CO$_2$ \citep{hu2021unveiling,wogan2024k218b} and NH$_3$ should be either removed by photolysis or dissolved in the oceans \citep{yu2021identify,hu2021unveiling}. The transmission spectrum of this scenario would have strong CO$_2$ features, which is also clearly ruled out by the transmission spectrum presented here (Figure~\ref{fig:lhs_selfmod}).

    Lastly, we considered an N$_2$-dominated atmosphere (with 10\% CO$_2$) and a CO$_2$-dominated atmosphere (with 10\% N$_2$) for the planet. With a surface pressure of 2.2 bar, we found that the surface temperatures of these two scenarios would be $\sim310$ and $\sim340$ K, consistent with liquid water. We caution that these temperatures are substantially higher than the surface temperatures predicted by 3D climate models \citep{cadieux2024lhsmass}. One possible reason for the discrepancy may be that the 3D models have a proper account of the ice-albedo effect. With the high mean molecular weight to subdue the transmission features, these models provide a reasonable fit to the spectrum (Figure \ref{fig:lhs_selfmod}). Notably, the models do not show significant absorption of water vapor as it has condensed out from the part of the atmosphere probed by transmission spectra, and the only spectral features that could be observed are CO$_2$ absorption. This is consistent with the models presented in \cite{cadieux2024lhsmass}. At $\sim4.2\ \mu$m, our models suggest a $\sim$20-ppm CO$_2$ absorption feature.
            
	\section{Discussion} \label{sec:discussion}

    \subsection{Excluding the H$_2$-rich atmosphere scenarios}

    The observations presented here were originally designed to detect an H$_2$-rich atmosphere with the possible presence of water vapor and other gases (e.g., methane and carbon dioxide). However, the transmission spectrum obtained makes an H$_2$-dominated atmosphere highly unlikely. Our atmospheric models show that, regardless of the atmospheric metallicity or size, an H$_2$-rich atmosphere on \lhs\ should result in detectable spectral features of either CH$_4$ or CO$_2$ (Figure \ref{fig:lhs_selfmod}). Because of the low temperature of the planet, H$_2$O is condensed out at a higher pressure than the region typically probed by transmission spectra. This eliminates the potentially confounding scenario of a highly metal-rich H$_2$ atmosphere \citep{benneke2024jwst}, because such a scenario would still result in detectable spectral features of CH$_4$ and CO$_2$ on \lhs.
            
    From the point of view of spectral retrievals, Table~\ref{tab:retrieval_results} shows that the data favor a high mean molecular weight atmosphere with the presence of water vapor. The data could also be marginally explained by an H$_2$-dominated atmosphere with high clouds ($\sim$10$^{2.8}$ Pa) and low water vapor mixing ratio ($\sim$10$^{-4}$) (Scenario 6, see Tables~\ref{tab:retrieval_values_phys} and \ref{tab:retrieval_values_chem} for detailed constraints). However, our atmospheric models show that the water clouds should extend to $\sim10^4$ Pa but not lower pressures. We have also checked for the condensation of NH$_3$ and the formation of NH$_4$SH clouds in the self-consistent atmospheric models shown in Figures \ref{fig:lhs_tp} and \ref{fig:lhs_selfmod} using the method of \cite{hu2021photochemistry}, and found that NH$_3$ should not condense, and the NH$_4$SH clouds, if forming, should only occur within a pressure scale height of the cold trap, having a minimal impact to the transmission spectra. Besides, such an H$_2$-dominated atmosphere should also have an appreciable abundance of CH$_4$ (if massive) or CO$_2$ (if small), and they would result in spectral features detectable in $3-5\ \mu$m. Therefore the H$_2$-dominated atmosphere with high clouds scenario is unlikely to apply to \lhs.

    One might ask if a high-altitude haze layer might create the flat spectrum. Given the low temperature of the planet, the plausible photochemical haze in a H$_2$-rich atmosphere includes sulfur haze from H$_2$S and hydrocarbon haze from CH$_4$. Atmospheric chemistry calculations indicated that the sulfur haze, if any, should be located at a similar pressure level as the water cloud and thus not impact the transmission spectrum \citep{hu2021photochemistry}. Hydrocarbon haze, on the other hand, can be produced from the photolysis of CH$_4$ in the upper atmosphere. However, this mechanism cannot explain the transmission spectrum of \lhs\ because, first, we do not detect any signals of CH$_4$ and second, such a haze layer would result in a slope in the transmission spectrum in the spectral range probed by JWST observations \citep{robinson2014titan,kawashima2019detectable,gao2020deflating}, which is not observed here. Lastly, we would like to point out that any massive H$_2$-rich atmosphere on \lhs\ should have abundant CH$_4$, which would cause a moderate temperature inversion in the middle atmosphere due to shortwave absorption (Figure \ref{fig:lhs_tp}). Such a stratified atmosphere would also facilitate the fall off of any large photochemical haze particles, giving rise to a spectral slope if any detectable photochemical haze is present at all.
            
            
            
    \subsection{Possible habitable water world} \label{sec:ww}

                \begin{figure}
                \centering
                \includegraphics[width=1\linewidth]{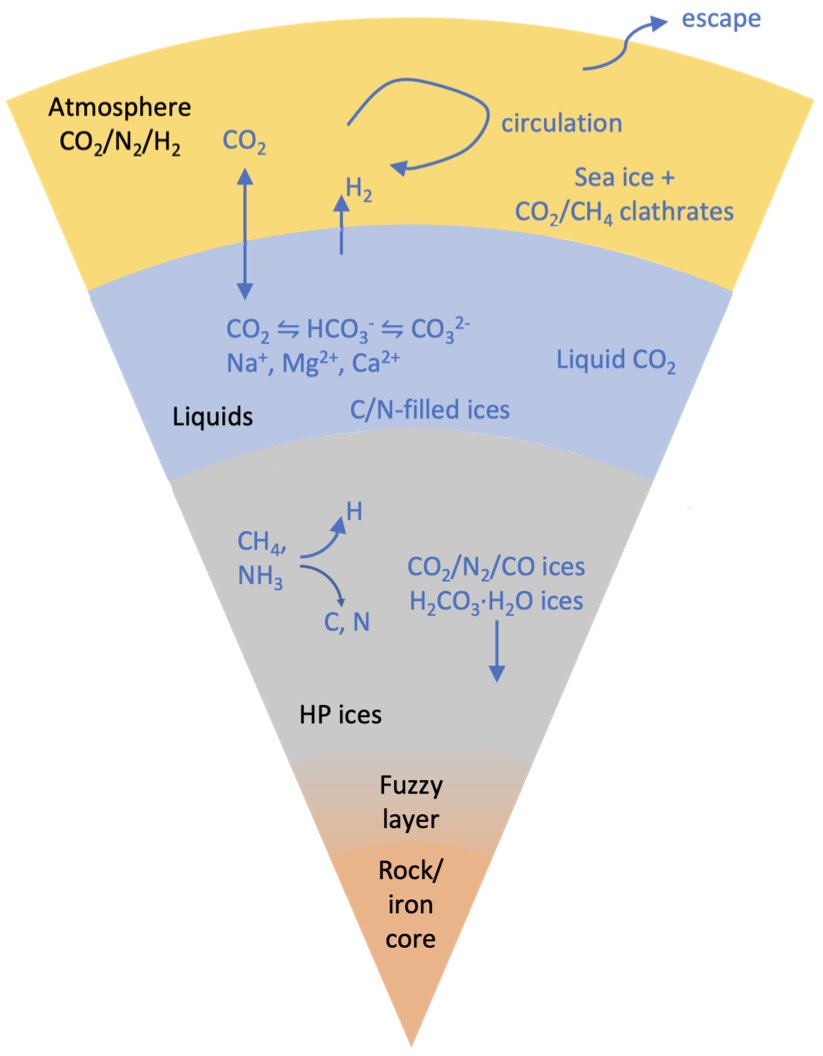}
                \caption{Potential processes that partition C and N species between the atmosphere, the ocean, and the HP ices in a cold water world like \lhs. Summarized based on \cite{levi2017abundance,ramirez2018ice,levi2019equation,marounina2020internal,vazan2022new,kovavcevic2022miscibility}.}
                \label{fig:ww}
            \end{figure}
        
    Without a massive H$_2$-rich atmosphere, the most plausible explanation for \lhs's low density is a water-rich envelope. Given the low irradiation and $\sim10\%$ water by mass \citep{cadieux2024lhsmass}, \lhs\ is likely to have a high-pressure (HP) ice mantle \citep[e.g.,][]{sotin2007mass,fu2009interior,zeng2014effect}, which itself may be partly or fully mixed with the rocky mantle underneath \citep[e.g.,][]{vazan2022new,kovavcevic2022miscibility}. Meanwhile, we could expect that C- and N-bearing ices were accreted together with H$_2$O ice when the planet formed \citep{oberg2011effects,schwarz2014effects}. With the accretional heat, those C- and N-bearing molecules would be thermally equilibrated with a reservoir of H$_2$O, resulting in predominantly CO$_2$ and N$_2$. Thus, the planet could reasonably have an N$_2$- or CO$_2$-dominated atmosphere. 
            
    The partitioning of C, N, and O molecules between the ice/rock mantle and the atmosphere controls the atmospheric size and composition \citep{levi2017abundance,marounina2020internal}. For example, depending on the planet's thermal evolution history, CO$_2$ can be stored in the interior as clathrate hydrates, liquids, and various types of ices (Figure~\ref{fig:ww}), which could give rise to a moderate-size atmosphere and a surface temperature conducive to liquid water. Alternatively, if the entirety of the planet's carbon and nitrogen presents as gas-phase N$_2$ and CO$_2$, this could result in a massive atmosphere and supercritical water layer \citep{levi2017abundance,marounina2020internal}. 
                
    It is therefore crucial to maintain the size of the N$_2$-CO$_2$ atmosphere for a habitable state to emerge. Silicate weathering as the key atmosphere/climate stabilizing mechanism for rocky planets is likely not applicable for \lhs, due to a lack of exposed landmasses and decoupling between the deep ocean and the atmosphere \citep{abbot2012indication}. Alternatively, the entrapment and depletion of CO$_2$ as clathrate-rich sea ice could help sustain a stable CO$_2$ atmospheric pressure suitable for liquid water \citep{ramirez2018ice}. This mechanism works because CO$_2$ clathrate hydrate is denser than water and is thermodynamically stable in the deep CO$_2$-saturated ocean. Therefore, the formation and sinking of CO$_2$ clathrates in the cooler part of the planet (e.g., the night side) could maintain a CO$_2$ atmosphere to $<\sim10$ bar. The workings of this mechanism on \lhs, which can be assumed to be tidally locked \citep{leconte2015asynchronous}, requires further studies.

    \subsection{Prospect for future observations}
    
    The N$_2$-CO$_2$ atmosphere scenarios result in small but potentially detectable spectral features of CO$_2$ absorption (Fig.~\ref{fig:lhs_selfmod}).
    Using \textsc{PandExo} \citep{batalha2017pandexo}, we find that an additional 9 transits with NIRSpec/G395M would be required to achieve a precision of 20 ppm at $\sim4.2\ \mu$m to detect and quantify the CO$_2$ absorption feature. This estimate is in line with those presented in \cite{cadieux2024lhsmass}. 

    Considering the rarity of \lhs\ transit events, a campaign of 9 additional transits would require a few years to complete. In addition, obtaining transit measurements using NIRISS/SOSS could help to disentangle the stellar heterogeneity component from the planetary atmospheric signature, benefiting from the continuous coverage from 0.6 to 2.8 $\mu$m. However, we do not yet understand how the stellar activity would change from visit to visit and this may prevent combining multiple NIRISS observations or using NIRISS observations to constrain NIRSpec/G395M observations directly. We note that two transit observations of \lhs\ have been taken using NIRISS/SOSS in December 2023 (Program ID: 6543, PI: C. Cadieux). Those observations could provide updated constraints on the stellar heterogeneity component and refinements to the transmission spectrum of \lhs\ at wavelengths $<3\ \mu$m. 
    
    \section{Conclusions} \label{sec:conclusion}

    In this paper, we present the data analysis of the two primary transit observations of \lhs\ performed by JWST using the NIRSpec instrument. The observations cover a spectral range between 1.7 and 5.2 $\mu$m. The transmission spectrum resulting from the JWST data shows a constant transit depth as a function of wavelength, indicating no apparent absorption features. Because an H$_2$-rich atmosphere on this planet would show strong transmission spectral features from CH$_4$ or CO$_2$, the spectrum presented here rules out an H$_2$-rich atmosphere on \lhs.
        
    We used a Bayesian retrieval framework, \texttt{ExoTR}, to interpret the transmission spectrum together with the stellar heterogeneity and found that the data favors an N$_2$-dominated atmosphere with H$_2$O and CO$_2$ over an H$_2$-dominated atmosphere with high clouds, or an H$_2$/N$_2$-dominated atmosphere without any molecular absorption, or a bare rock scenario by $\sim3\sigma$ confidence. The H$_2$-dominated atmosphere with high clouds or photochemical haze is also unlikely because the expected water cloud is deep in the atmosphere (at $>\sim10^4$ Pa) and the spectrum does not show any signals of CH$_4$ which is the feedstock to form photochemical haze.

    The observation and analysis presented here effectively leave a water-dominated layer as the only plausible explanation for the low density of the planet. Also, the water world cannot have a small H$_2$-dominated atmosphere, i.e., not a ``hycean world.'' Rather, based on planetary evolution considerations, we suggest that an N$_2$- or CO$_2$-dominated atmosphere is most likely and consistent with the transmission spectrum measured here. Our climate model indicates that a moderate-size ($\sim2$ bar) N$_2$- or CO$_2$-dominated atmosphere could maintain a global mean surface temperature above the freezing point of water. If the planet evolves to or has the climate-stabilizing mechanism to maintain such a moderate-size N$_2$- or CO$_2$-dominated atmosphere, the planet may be a habitable water world. 


    \lhs\ is thus a unique planet that provides the rare opportunity to observe and characterize a temperate, potentially habitable water worlds. We estimated that 9 additional transits may be required to detect CO$_2$ in the high mean molecular weight atmosphere on this planet, and this could be completed in $\sim3$ JWST cycles. With the existence of an atmosphere on TRAPPIST-1 planets called into question \citep{dong2018atmospheric,greene2023thermal,zieba2023no}, \lhs\ may well present the best current opportunity to detect and characterize a habitable world in our interstellar neighborhood.


    \section*{Data availability}
    All the {\it JWST} data used in this paper can be found in MAST: \dataset[10.17909/r627-v590]{http://dx.doi.org/10.17909/r627-v590}. Raw data will be publicly available after the proprietary period (i.e., August 2024).
 
	\section*{Acknowledgments}
	We thank Amit Levi, Leslie Rogers, Ramses Ramirez, and Allona Vazan for helpful discussion on the evolution of water worlds. This research is based on observations with the NASA/ESA/CSA James Webb Space Telescope obtained \dataset[the dataset]{http://dx.doi.org/10.17909/r627-v590} at the Space Telescope Science Institute, which is operated by the Association of Universities for Research in Astronomy, Incorporated, under NASA contract NAS5-03127. Support for Program number 2334 was provided through a grant from the STScI under NASA contract NAS5-03127. This research was carried out at the Jet Propulsion Laboratory, California Institute of Technology, under a contract with the National Aeronautics and Space Administration (80NM0018D0004). The High Performance Computing resources used in this investigation were provided by funding from the JPL Information and Technology Solutions Directorate.
 
    \section*{Software}
    \noindent \exotr\ (\href{https://github.com/MDamiano/ExoTR}{GitHub}),
    \textsc{Numpy} \citep{oliphant2015numpy}, \textsc{Scipy} \citep{virtanen2020scipy}, \textsc{Astropy} \citep{astropy2013,astropy2018,astropy2022}, \textsc{scikit-bio} \citep{skbio2020}, \textsc{Matplotlib} \citep{hunter2007matplotlib}, \textsc{MultiNest} \citep{feroz2009multinest,buchner2014multinest}, \textsc{mpi4py} \citep{dalcin2021mpi4py}, \textsc{spotrod} \citep{beky2014spotrod}, \textsc{Eureka!} \citep{bell2022eureka}, \textsc{ExoTiC-LD} \citep{grant2022exotic}, \textsc{batman} \citep{kreidberg2015batman}, \textsc{PandExo} \citep{batalha2017pandexo}. 

    \newpage
	
	{	\small
		\bibliographystyle{apj}
		\bibliography{bib.bib}
	}

    \appendix

    \section{Scenarios retrieval results} \label{sec:appendixA}
        Tables \ref{tab:retrieval_values_phys} and \ref{tab:retrieval_values_chem} contain the results of the retrieval analyses performed on the scenarios listed in Table~\ref{tab:retrieval_results}.

        \begin{deluxetable}{ccccccccc} [!h]
            \tablecaption{Retrieval results for the scenarios presented in Table~\ref{tab:retrieval_results}. The offsets are defined relative to the NIRSpec/G325H-NRS1 dataset. Therefore, $off_1$ is the offset between NRS1 and NRS2 within the G235H grating, $off_2$ is the offset between the G325H-NRS1 and G395H-NRS1, and $off_3$ is the offset between the G325H-NRS1 and G395H-NRS2. \label{tab:retrieval_values_phys}}
            \tablehead{\textbf{Scenario} & $off_1$ & $off_2$ & $off_3$ & R$_p$ [R$_{\oplus}$] & Log(P$_{top}$) & $\delta$ & T$_{het}$ & T$_{phot}$}
            \startdata
            \textbf{1.} & -0.73$^{+8.99}_{-8.78}$ & -52.89$^{+8.89}_{-9.02}$ & -59.20$^{+10.79}_{-10.53}$ & 1.730$\pm$0.003 & 6.48$^{+1.68}_{-1.94}$ & $-$ & $-$ & $-$ \\
            \textbf{2.} & 0.23$^{+7.78}_{-9.07}$ & -51.52$^{+9.05}_{-9.57}$ & -66.35$^{+10.51}_{-10.98}$ & 1.732$\pm$0.002 & 6.49$^{+1.66}_{-1.85}$ & $-$ & $-$ & $-$\\
            \textbf{3.} & 2.42$^{+9.37}_{-7.95}$ & -51.57$^{+9.22}_{-8.81}$ & -60.85$^{+11.03}_{-9.95}$ & 1.728$\pm$0.002 & 5.93$^{+1.93}_{-1.91}$ & $-$ & $-$ & $-$ \\
            \textbf{4.} & -2.10$^{+8.32}_{-8.34}$ & -59.88$^{+10.52}_{-11.58}$ & -69.01$^{+12.98}_{-14.10}$ & 1.725$\pm$0.006 & 6.37$^{+1.69}_{-1.97}$ & 0.02$^{+0.03}_{-0.01}$ & 2641$^{+410}_{-517}$ & 3098$\pm$45\\
            \textbf{5.} & 1.39$^{+8.23}_{-8.76}$ & -57.26$^{+10.46}_{-10.73}$ & -68.30$^{+13.33}_{-13.98}$ & 1.724$\pm$0.004 & 5.91$^{+1.99}_{-2.15}$ & 0.03$^{+0.05}_{-0.02}$ & 2941$^{+168}_{-325}$ & 3097$\pm$41 \\
            \textbf{6.} & 4.25$^{+9.09}_{-9.23}$ & -57.14$^{+9.72}_{-8.08}$ & -64.45$^{+8.89}_{-9.82}$ & 1.657$\pm$0.007 & 2.82$^{+0.45}_{-0.58}$ & $-$ & $-$ & $-$ \\
            \textbf{7.} & -5.09$^{+9.07}_{-8.82}$ & -54.74$^{+9.87}_{-9.82}$ & -60.72$^{+10.71}_{-11.13}$ & 1.732$\pm$0.003 & 3.59$^{+2.99}_{-2.46}$ & $-$ & $-$ & $-$\\
            \textbf{8.} & -4.15$^{+8.93}_{-9.22}$ & -53.53$^{+9.81}_{-10.06}$ & -59.88$^{+10.62}_{-10.36}$ & 1.737$\pm$0.001 & $-$ & $-$ & $-$ & $-$ \\
            \textbf{9.} & -4.16$^{+8.84}_{-8.97}$ & -53.81$^{+8.91}_{-9.96}$ & -60.03$^{+10.53}_{-10.76}$ & 1.641$\pm$0.013 & 1.59$^{+1.04}_{-1.06}$ & $-$ & $-$ & $-$ \\
            \enddata
        \end{deluxetable}

        \begin{deluxetable}{ccccccccc}[!h]
            \tablecaption{Retrieval results on the atmospheric abundances for the scenarios presented in Table~\ref{tab:retrieval_results}. \label{tab:retrieval_values_chem}}
            \tablehead{\textbf{Scenario} & Log(H$_2$O) & Log(CH$_4$) & Log(NH$_3$) & Log(CO) & Log(CO$_2$) & Log(N$_2$) & Log(H$_2$) & $\mu$ (derived)}
            \startdata
            \textbf{1.} & -0.34$^{+0.33}_{-3.38}$ & $-$ & $-$ & $-$ & -2.32$^{+1.51}_{-3.09}$ & -0.84$^{+0.84}_{-3.25}$ & $-$ & 25.45$^{+2.65}_{-7.19}$\\
            \textbf{2.} & -2.19$^{+1.61}_{-2.13}$ & $-$ & $-$ & $-$ & $-$ & -0.01$^{+0.01}_{-0.13}$ & $-$ & 27.95$^{+0.06}_{-2.55}$\\
            \textbf{3.} & -0.05$^{+0.05}_{-1.42}$ & -6.83$^{+1.78}_{-2.03}$ & -4.58$^{+3.04}_{-3.26}$ & -3.18$^{+2.58}_{-3.22}$ & -4.32$^{+2.98}_{-3.77}$ & -3.91$^{+2.78}_{-3.33}$ & -5.16$^{+3.18}_{-3.91}$ & 18.50$^{+8.97}_{-0.52}$\\
            \textbf{4.} & -1.31$^{+1.29}_{-3.03}$ & $-$ & $-$ & $-$ & -2.71$^{+2.02}_{-3.59}$ & -0.15$^{+0.15}_{-2.77}$ & $-$ & 27.98$^{+0.79}_{-9.17}$\\
            \textbf{5.} & -0.27$^{+0.26}_{-2.21}$ & -6.53$^{+1.97}_{-2.26}$ & -4.55$^{+3.12}_{3.09}$ & -3.00$^{+2.54}_{-3.23}$ & -3.43$^{+2.81}_{-4.22}$ & -3.19$^{+2.64}_{-3.60}$ & -5.14$^{+3.45}_{-4.67}$ & 20.42$^{+7.73}_{-2.43}$\\
            \textbf{6.} & -4.20$^{+0.93}_{-0.59}$ & $-$ & $-$ & $-$ & $-$ & $-$ & -0.01$^{+0.01}_{-0.01}$ & 2.02$\pm$0.01 \\
            \textbf{7.} & $-$ & $-$ & $-$ & $-$ & $-$ & 0.00 & $-$ & 28.00\\
            \textbf{8.} & $-$ & $-$ & $-$ & $-$ & $-$ & $-$ & $-$ & $-$\\
            \textbf{9.} & $-$ & $-$ & $-$ & $-$ & $-$ & $-$ & 0.00 & 2.02\\
            \enddata
        \end{deluxetable}

    \newpage

    \section{\lhs\ JWST/NIRSpec 1D transmission spectrum \label{sec:appendixC}}
        The 1D transmission spectrum, calculated after reducing and analyzing the data and shown throughout the manuscript, is reported in Table \ref{tab:spec235} and \ref{tab:spec395}. NOTE - The offsets are not included in Table \ref{tab:spec235} and \ref{tab:spec395} and if needed they can be adopted from Table \ref{tab:retrieval_values_phys}.

        \begin{deluxetable}{ccc|ccc}[!h]
            \tablecaption{\lhs\ NIRSpec/G235H 1D transmission spectrum (R=65). \label{tab:spec235}}
            \tablehead{\multicolumn{3}{c}{\textbf{G235H-NRS1}} & \multicolumn{3}{c}{\textbf{G235H-NRS2}}}
            \startdata
            \textbf{Wavelength bin [$\mu$m]} & \textbf{(R$_p$/R$_s$)$^2$ [ppm]} & \textbf{$\Delta$(R$_p$/R$_s$)$^2$ [ppm]} & \textbf{Wavelength bin [$\mu$m]} & \textbf{(R$_p$/R$_s$)$^2$ [ppm]} & \textbf{$\Delta$(R$_p$/R$_s$)$^2$ [ppm]}\\
            \hline
            1.6652 $-$ 1.6908 & 5378.13 & 25.98 & 2.2662 $-$ 2.3011 & 5429.56 & 22.78 \\
            1.6908 $-$ 1.7168 & 5422.31 & 26.67 & 2.3011 $-$ 2.3365 & 5378.50 & 24.18 \\
            1.7168 $-$ 1.7433 & 5396.81 & 28.85 & 2.3365 $-$ 2.3724 & 5438.49 & 24.43 \\ 
            1.7433 $-$ 1.7701 & 5452.86 & 28.43 & 2.3724 $-$ 2.4089 & 5499.62 & 25.45 \\
            1.7701 $-$ 1.7973 & 5327.54 & 28.39 & 2.4089 $-$ 2.4460 & 5427.84 & 26.17 \\ 
            1.7973 $-$ 1.8250 & 5516.30 & 27.66 & 2.4460 $-$ 2.4836 & 5439.00 & 27.50 \\
            1.8250 $-$ 1.8530 & 5453.87 & 26.95 & 2.4836 $-$ 2.5218 & 5433.92 & 28.85 \\ 
            1.8530 $-$ 1.8815 & 5454.42 & 26.62 & 2.5218 $-$ 2.5606 & 5432.32 & 29.76 \\ 
            1.8815 $-$ 1.9105 & 5498.23 & 26.57 & 2.5606 $-$ 2.6000 & 5430.87 & 30.19 \\ 
            1.9105 $-$ 1.9399 & 5466.28 & 25.92 & 2.6000 $-$ 2.6400 & 5534.57 & 30.89 \\ 
            1.9399 $-$ 1.9697 & 5422.28 & 25.35 & 2.6400 $-$ 2.6806 & 5494.14 & 30.24 \\ 
            1.9697 $-$ 2.0000 & 5446.42 & 25.46 & 2.6806 $-$ 2.7219 & 5461.31 & 32.24 \\ 
            2.0000 $-$ 2.0308 & 5473.55 & 25.17 & 2.7219 $-$ 2.7637 & 5431.83 & 35.23 \\
            2.0308 $-$ 2.0620 & 5470.70 & 24.79 & 2.7637 $-$ 2.8063 & 5382.22 & 36.90 \\ 
            2.0620 $-$ 2.0938 & 5449.93 & 25.15 & 2.8063 $-$ 2.8494 & 5463.46 & 37.26 \\ 
            2.0938 $-$ 2.1260 & 5381.30 & 24.33 & 2.8494 $-$ 2.8933 & 5484.73 & 37.83 \\ 
            2.1260 $-$ 2.1587 & 5456.75 & 23.78 & 2.8933 $-$ 2.9378 & 5403.74 & 39.06 \\
            2.1587 $-$ 2.1919 & 5457.52 & 24.02 & 2.9378 $-$ 2.9830 & 5415.41 & 38.45 \\
            2.1919 $-$ 2.2256 & 5437.82 & 41.89 & 2.9830 $-$ 3.0289 & 5463.53 & 40.23 \\ 
                              &         &       & 3.0289 $-$ 3.0755 & 5518.91 & 43.33 \\
            \enddata
        \end{deluxetable}

        \begin{deluxetable}{ccc|ccc}[!h]
            \tablecaption{\lhs\ NIRSpec/G395H 1D transmission spectrum (R=65). \label{tab:spec395}}
            \tablehead{\multicolumn{3}{c}{\textbf{G395H-NRS1}} & \multicolumn{3}{c}{\textbf{G395H-NRS2}}}
            \startdata
            \textbf{Wavelength bin [$\mu$m]} & \textbf{(R$_p$/R$_s$)$^2$ [ppm]} & \textbf{$\Delta$(R$_p$/R$_s$)$^2$ [ppm]} & \textbf{Wavelength bin [$\mu$m]} & \textbf{(R$_p$/R$_s$)$^2$ [ppm]} & \textbf{$\Delta$(R$_p$/R$_s$)$^2$ [ppm]}\\
            \hline
                2.8799 $-$ 2.9242 & 5509.11 & 38.77 & 3.8241 $-$ 3.8829 & 5492.83 & 29.44 \\ 
                2.9242 $-$ 2.9692 & 5498.85 & 36.04 & 3.8829 $-$ 3.9426 & 5478.97 & 30.07 \\ 
                2.9692 $-$ 3.0148 & 5443.82 & 34.16 & 3.9426 $-$ 4.0033 & 5464.45 & 31.01 \\ 
                3.0148 $-$ 3.0612 & 5467.37 & 34.27 & 4.0033 $-$ 4.0649 & 5505.28 & 31.96 \\ 
                3.0612 $-$ 3.1083 & 5493.88 & 31.88 & 4.0649 $-$ 4.1274 & 5550.53 & 33.25 \\ 
                3.1083 $-$ 3.1561 & 5481.44 & 31.34 & 4.1274 $-$ 4.1909 & 5499.78 & 34.27 \\ 
                3.1561 $-$ 3.2047 & 5546.91 & 31.41 & 4.1909 $-$ 4.2554 & 5437.88 & 34.99 \\ 
                3.2047 $-$ 3.2540 & 5487.67 & 30.90 & 4.2554 $-$ 4.3209 & 5473.49 & 40.13 \\ 
                3.2540 $-$ 3.3041 & 5491.85 & 30.05 & 4.3209 $-$ 4.3873 & 5507.15 & 38.93 \\ 
                3.3041 $-$ 3.3549 & 5494.97 & 29.73 & 4.3873 $-$ 4.4548 & 5569.35 & 41.51 \\ 
                3.3549 $-$ 3.4065 & 5501.71 & 29.31 & 4.4548 $-$ 4.5234 & 5589.28 & 42.94 \\ 
                3.4065 $-$ 3.4589 & 5483.22 & 29.12 & 4.5234 $-$ 4.5930 & 5488.92 & 44.56 \\ 
                3.4589 $-$ 3.5121 & 5478.35 & 29.03 & 4.5930 $-$ 4.6636 & 5503.88 & 45.66 \\ 
                3.5121 $-$ 3.5662 & 5517.34 & 28.91 & 4.6636 $-$ 4.7354 & 5567.58 & 47.10 \\ 
                3.5662 $-$ 3.6210 & 5451.03 & 28.41 & 4.7354 $-$ 4.8082 & 5470.57 & 48.58 \\ 
                3.6210 $-$ 3.6767 & 5542.96 & 28.97 & 4.8082 $-$ 4.8822 & 5462.63 & 50.78 \\ 
                3.6767 $-$ 3.7333 & 5521.20 & 34.13 & 4.8822 $-$ 4.9573 & 5531.78 & 52.95 \\ 
                                  &         &       & 4.9573 $-$ 5.0336 & 5452.13 & 55.20 \\
                                  &         &       & 5.0336 $-$ 5.1110 & 5537.30 & 55.82 \\
                                  &         &       & 5.1110 $-$ 5.1896 & 5442.32 & 80.72 \\
            \enddata
        \end{deluxetable}

    \newpage

    \section{Joint fit of ground-based, HST, and JWST data} \label{sec:appendixB}
        
        Given its intriguing physical parameters, \lhs\ has been the object of several observation campaigns to unveil the atmospheric composition and assess the nature of the planet. 
        HST initially observed the planet in January and December of 2017 (program ID: 14888, PI: Dittmann, J.) for an initial reconnaissance of the planet's transmission spectrum between $\sim$1.1 and $\sim$1.6 $\mu$m. The resulting 1D transmission spectrum suggests strong spectral modulations that may be attributed to H$_2$O. If the spectral modulation truly has a planetary origin, the planet should have an H$_2$-dominated atmosphere \citep{edwards2020lhshst}. Another campaign was conducted between 2017 and 2018 to observe the planet in transmission spectroscopy in the visible band with the Magellan telescope at Las Campanas Observatory. The obtained spectrum revealed a strong linear trend due to the stellar heterogeneity \citep{diamond2020lhsoptical}. 

\begin{figure*}[h!]
                \centering
                \includegraphics[width=\linewidth]{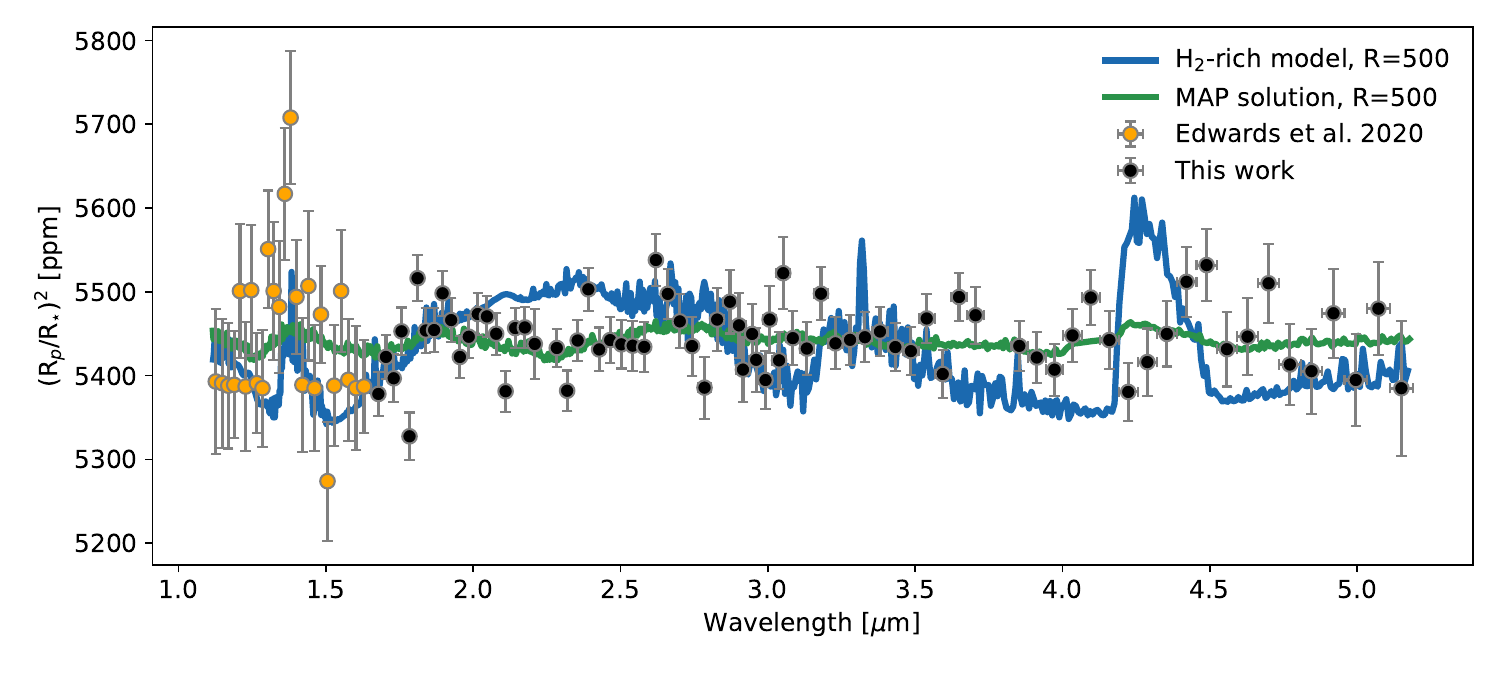}
                \caption{The transmission spectrum obtained in this work compared with the HST data adopted from \cite{edwards2020lhshst}. The green model, i.e., maximum a posteriori (MAP) solution of scenario 4 in Table \ref{tab:retrieval_results}, extended to the HST data wavelength. The blue model is the best fit model for the HST data (an H$_2$-dominated atmosphere with water vapour absorption) with a CH$_4$ and CO$_2$ 10$^{-6}$ VMR injected extended to the wavelength range covered by the JWST data. The JWST dataset is incompatible with the blue model. \label{fig:h2_model}}
            \end{figure*}

        The HST and JWST datasets do not appear to be consistent with each other. Fig.~\ref{fig:h2_model} compares the solution of the spectral retrieval performed on the HST observation with the JWST data, and it shows that the data and the model are incompatible. In particular, the lack of the H$_2$-H$_2$ CIA spectral feature is clear between 2 and 3 $\mu$m.

\begin{figure*}[!h]
            \centering
            \includegraphics[width=\linewidth]{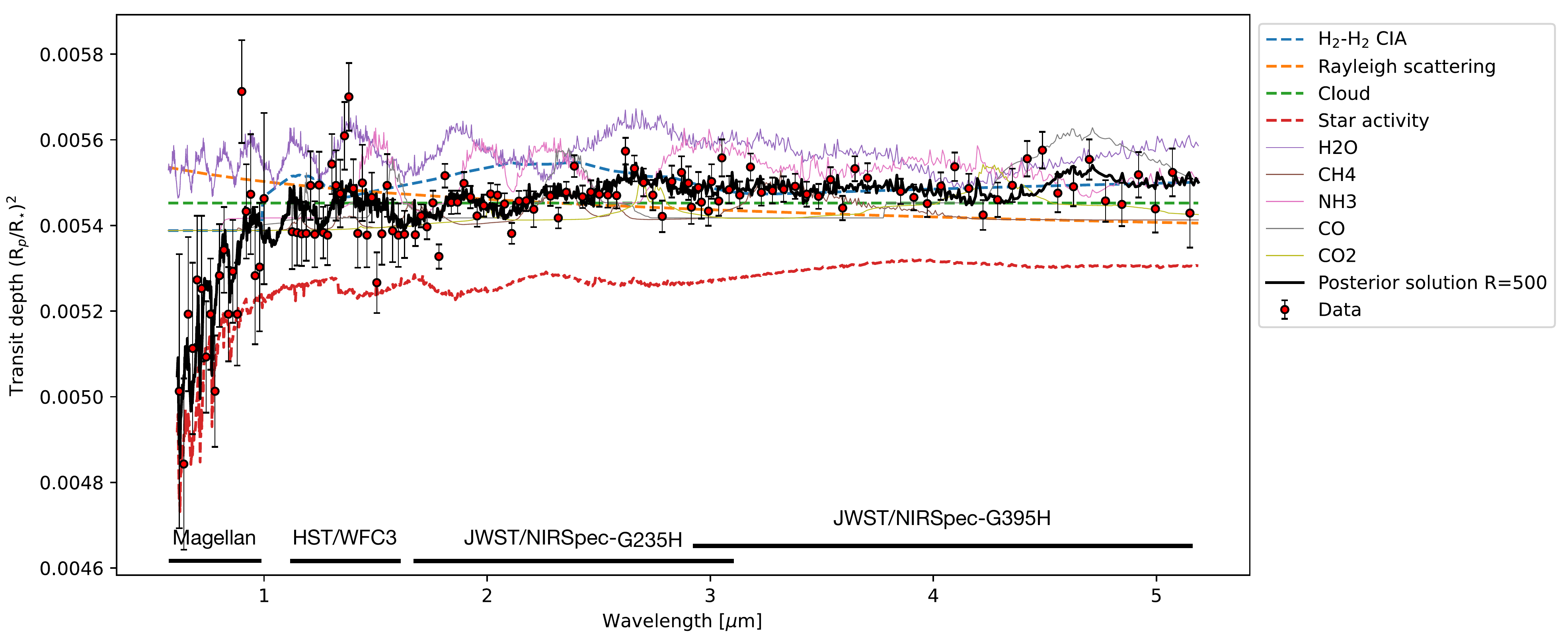}
            \caption{Available datasets for \lhs. The maximum a posteriori solution from the Bayesian retrieval process is shown in black. The individual contribution from the spectral components is also shown. The data is plotted by taking into consideration the offsets value calculated by the Bayesian analysis. The offsets value are reported in the posterior distribution functions in Fig.~\ref{fig:all_spec_post}. HST data are adopted from \citep{edwards2020lhshst} and the Magellan data are adopted from \citep{diamond2020lhsoptical}. \label{fig:all_spec}}
        \end{figure*}

\begin{figure*}[!t]
            \centering
            \includegraphics[width=\linewidth]{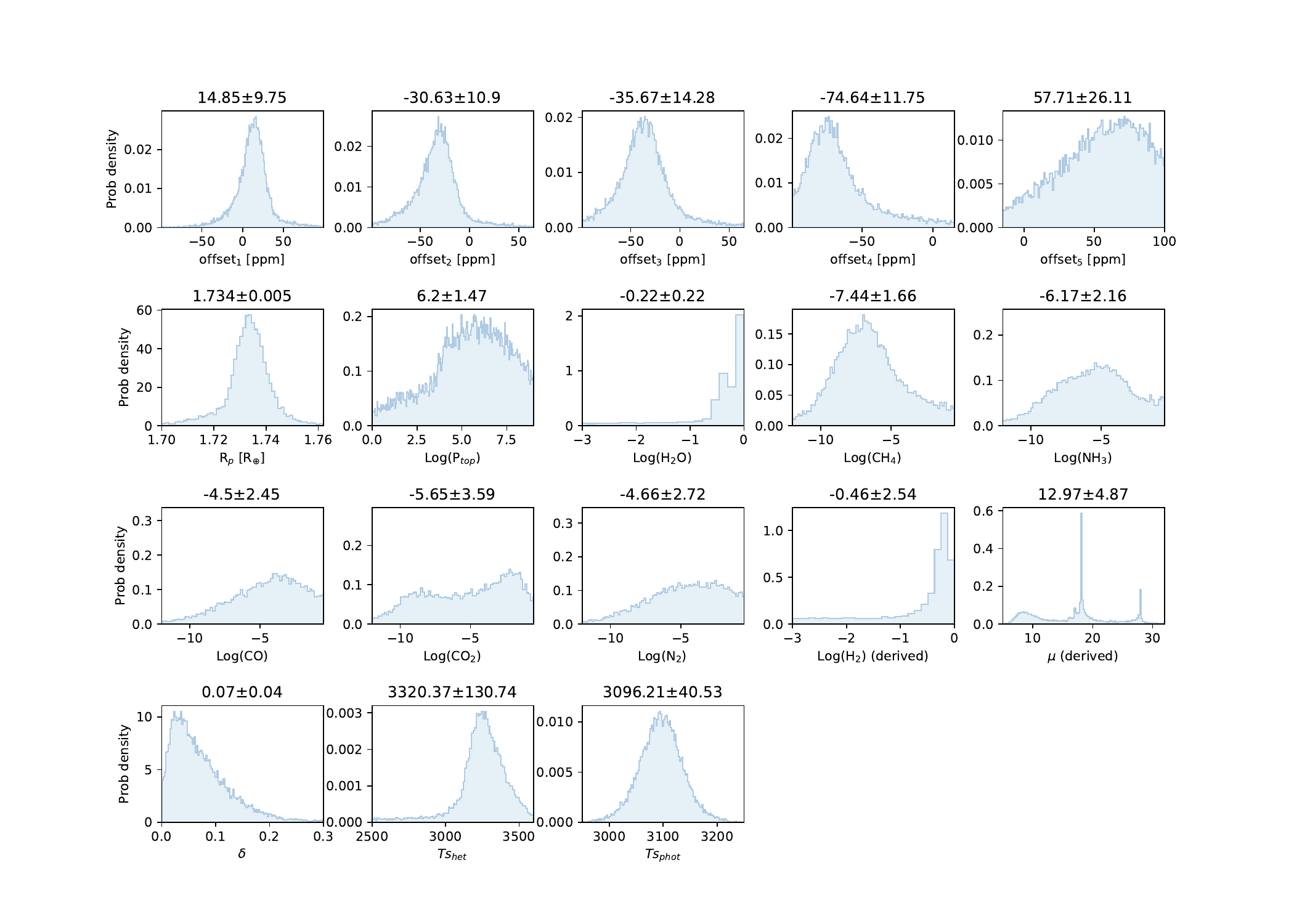}
            \caption{Posterior distribution functions of the Bayesian analysis of all the \lhs\ datasets combined. The posterior distribution functions suggest an H$_2$O-rich atmosphere with a non-negligible presence of hydrogen. The offsets are defined as follows: $offset_{1}$ is related with the G235H - NRS2, and $offset_{2}$ and $offset_{3}$ are instead relative to the G395H - NRS1 and NRS2 respectively. $offset_{4}$ is between the HST data \citep{edwards2020lhshst} and G325H-NRS1 dataset and $offset_{5}$ is between the Magellan dataset \citep{diamond2020lhsoptical} and the G325H-NRS1 data. The numeric values above each panel indicate the median and $1\sigma$ uncertainty of the distribution. \label{fig:all_spec_post}}
        \end{figure*}

        Under the assumption that the stellar heterogeity signal remains constant between epoches (which is likely not true), we performed a joint fit of all the datasets aforementioned together with the the JWST data presented in this work (Figs.~\ref{fig:all_spec} and \ref{fig:all_spec_post}) to fit the stellar heterogeity component and assess whether or not a high mean molecular weight atmosphere is still a statistically preferred by the data. 
        The visible-wavelength dataset helps primarily to fit the stellar heterogeneity contribution to the transmission spectrum. The stellar heterogeneity also contributes at longer wavelengths, but not significantly beyond 3 $\mu$m. According to the resulting posterior distribution functions in Fig.~\ref{fig:all_spec_post}, the combined data do not favor an H$_2$-dominated atmosphere, but rather an H$_2$O-dominated atmosphere, potentially with some amounts of H$_2$. This is largely consistent with the results presented in Sec.~\ref{sec:result}, where a high mean molecular weight atmosphere is preferred. The non negligible amount of H$_2$ mainly comes form the fitting of the HST data. In that dataset, the spectral feature around 1.3$\mu$m is significant. The difference in (R$_p$/R$_{\star}$)$^2$ between the peak at 1.38 $\mu$m and the baseline is approximately 200 ppm. The pressure scale height of an H$_2$-dominated atmosphere around \lhs\ is $\sim$35 km, corresponding to 38 ppm. Therefore, the potential spectral variation would correspond to approximately 5 pressure scale heights in an H$_2$-dominated atmosphere. However, the JWST data, as explained in Sections~\ref{sec:result} and \ref{sec:discussion}, do not show such large spectral features. It is entirely possible that the HST dataset suffers from a particularly intense episode of stellar heterogeneity.

\end{document}